\documentclass[final,5p,times,twocolumn,authoryear]{elsarticle}

\usepackage{epsfig}
\usepackage{amssymb}
\usepackage{lipsum}
\usepackage{orcidlink}
\usepackage{graphicx}
\usepackage{natbib}

\usepackage{rotating}

\journal{New Astronomy}

\begin{document}

\begin{frontmatter}



\title{High Power Accretion in Massive Binary Systems and the Impact of Metallicity}


\author{Bhawna Mukhija$^{a}$ \orcidlink{0009-0007-1450-6490} and Amit Kashi$^{a,b}$ \orcidlink{0000-0002-7840-0181}\\}
\affiliation[]{orgnaization={ Department of Physics, Ariel University, Ariel, 4070000} \\$^{b}$ Astrophysics, Geophysics and Space Science (AGASS) Center, Ariel University, Ariel, 4070000, Israel}



\begin{abstract}

During a giant eruption of a very massive star in the binary system, the companion star can accrete a large amount of mass that can change its properties and potentially its subsequent evolution. The effect depends on the companion mass, metallicity, the amount of mass it accreted, orbital parameters and other parameters.
We simulate individual companion stars assuming they undergo such accretion events.
We study the envelope properties of 20 $\rm M_\odot$ and 30 $\rm M_\odot$ single massive stars at different matallicities ($Z= 0.02$, $Z=0.008$ and $Z=0.004$) during accretion at different rates, from $\rm 10^{-5}$ to $\rm 10^{-2}~M_\odot\,yr^{-1}$.
For the lower accretion rates we simulate, the stars remains hot, while at higher accretion rates, it becomes cooler and inflates. This behavior is observed in both stars but occurs at different accretion rates. Higher metallicity stars exhibit greater variations in accretion luminosity for the same accretion rate and stellar mass compared to lower metallicity stars. While higher metallicity stars typically have larger stellar envelopes, suggesting smaller variations in luminosity at Galactic metallicity compared to the LMC and SMC, our results show the opposite. 
\end{abstract}


\end{frontmatter}




\section{Introduction}
\label{1}
For a long time, it was thought that main sequence (MS) stars are unable to accrete mass at high rates ($\rm \gtrsim 10^{-3}~M_{\odot}~yr^{-1}$) \citep[e.g,][]{1971MNRAS.154..141H,1981ARA&A..19..137P, 1991ApJ...370..709H}. It was later suggested that MS stars could sustain high accretion rates ( $\rm \approx 10^{-3}~M_{\odot}~yr^{-1}$) in specific scenarios, particularly in the context of massive binary systems \citep[e.g.,][]{2001MNRAS.325..584S, 2010ApJ...723..602K}. However, observations also suggest that high mass accretion processes are possible and are now understood to be powered by super-Eddington accretion of matter onto stellar-mass compact objects \citep[e.g.,][]{2014Natur.514..202B, 2020A&A...642A.174M, 2023arXiv230200006P}.
Companion stars in the binary system expand if the infall of the material is too fast, and if the expansion is significant enough for the accretor also to fill its Roche lobe, a common envelope is likely to form \citep[e.g.,][]{1977A&A....54..539K, 1977PASJ...29..249N, 2013A&ARv..21...59I, 2017MNRAS.471.4839S, 2021A&A...650A.107M}. With further expansion, material is expected to leave the binary system through the second Lagrange point. This process can lead to the merger of the two stars, as the expelled material removes a substantial amount of angular momentum from the system \citep[e.g,][]{1971ARA&A...9..183P, 2003astro.ph..2232L,2017MNRAS.469.2441L,2011A&A...528A.114T,2014ApJ...786...39N, 2024arXiv240408615S}.

\citet{1991A&A...241..419P} found that the accretor does not undergo significant expansion as long as its accretion rate is less than ten times its thermal timescale accretion rate. This method is widely used in various binary population studies \citep[e.g.,][]{2002MNRAS.329..897H,2014ApJ...796...37S,2016ApJ...833..108S,2018MNRAS.481.4009V,2022ApJ...931...17V}, where the mass gain of the accretor is typically constrained by the thermal timescale accretion rate when the mass transfer rate surpasses this threshold \citep[e.g.,][]{1996A&A...309..179P, 2012A&A...546A..70T}.

\citet{2024arXiv240703182B} investigated how massive MS stars can accrete mass at very high rates without experiencing significant expansion. They proposed a mechanism in which the accreting star loses a substantial fraction of the incoming material from its outer layers via powerful jets launched by an accretion disk. These jets expel the high-entropy outer layers, stabilizing the star's structure and preventing excessive swelling. They linked this mechanism to several astrophysical phenomena, including intermediate luminosity optical transients \citep[e.g.,][]{2009ApJ...699.1850B, 2016RAA....16...99K, 2019PASP..131k8002M}, the grazing envelope evolution \citep[e.g.,][]{2020Galax...8...26S, 2023MNRAS.524L..94S}, and the 1837-1856 Great Eruption of Eta Carina \citep[e.g.,][]{2001MNRAS.325..584S, 2010ApJ...709L..11K}. \cite{Scolnic_2025} simulated mass accretion at very high rates onto massive MS stars combined with mass loss and found that they can accrete up to $10\%$ of their mass without expending much if the mass is being removed by jets.

Metallicity is also known to affect the evolution of a massive star significantly; the lower the metallicity, the more compact the star, especially during the post-MS evolution \citep[e.g.,][]{1982ApJS...49..447B,1991A&A...245..548B}. This difference can be more pronounced for stars with masses $\rm > 50~M_{\odot}$ \citep{2017A&A...597A..71S}. When the stellar evolution tracks are compared across different metallicities, massive stars exhibit a relatively compact size during their rapid Hertzsprung-Gap (HG) expansion phase, as was found for a large range of metallicities ($Z=0.0088, 0.0047, 0.0021$ by \citealt{2011A&A...530A.115B}; $Z=0.004$ by \citealt{2019A&A...627A..24G}; $Z=0.0034$ by \citealt{2019A&A...625A.132S}; $Z=0.002$ by  \citealt{2013A&A...558A.103G}; $Z=0$ ( pop III stars) by \citealt{2001A&A...371..152M}).
Such stars regain thermal equilibrium and start the core-helium burning phase as B- or A-type blue supergiants (BSGs $R \approx \rm 100~R_{\odot}$) and expand more to become red supergiants (RSGs $ R \approx \rm 1000~R_{\odot}$) during their final stages of core-helium burning. 

\citet{2015A&A...580A..20S} showed that the Eddington limit is rarely reached at the stellar surface, though stars more massive than $\rm 40\, M_{\odot}$ often locally exceed it within their envelopes, particularly in partial ionization zones. This local excess leads to strong structural changes such as envelope inflation, convection, and density inversions. All models with luminosities above $\rm 4 \times 10^{5}\, L_{\odot}$ show significant inflation, suggesting a link to S Doradus variability and major LBV eruptions. Thus, stars above $\rm 40\, M_{\odot}$, especially in the LMC, are expected to reach or exceed the Eddington limit internally. However, when the Eddington limit is defined in the stellar interior \citep{1997ASPC..120.....N}, \citet{2017A&A...597A..71S}, it is found that MS models with LMC composition reach and exceed the Eddington limit at masses above $\rm 40~M_{\odot}$, resulting in a process called envelope inflation. Where the star expands its envelope to dilute its structure rather than developing a strong outflow \citep{1999PASJ...51..417I}.
\citet{2016MNRAS.458.1214Q} explored the dynamics of winds driven by energy deposition near the surface of massive stars. These winds are relevant to events where super-Eddington luminosities are achieved, such as in the luminous blue variables (LBVs), Type IIn supernovae progenitors, classical novae, and X-ray bursts. They developed both analytic and numerical models to describe how near-surface heating, which arises from unstable fusion, wave heating, or binary interactions, drives strong stellar winds.

In \citet{2024ApJ...974..124M}, we studied the effects of a Giant Eruption (GE) on a massive star by creating an artificial eruption. This involved imposing a high mass loss rate of $\rm 0.15~M_{\odot}~yr^{-1} $ on the star's outer layers over a brief period. During the eruption, the star's luminosity decreased while its temperature increased, reflecting the significant impact of the mass loss process on its stellar characteristics and evolutionary trajectory on the HR diagram. Following the eruption, the study modeled the star's recovery phase, during which mass loss ceased, and the star gradually returned to equilibrium. During this phase, the star re-established its structure and moved toward the cooler region of the HR diagram, regaining stability and its pre-eruption characteristics.

Accretion onto a companion star in a massive binary system with an LBV primary was proposed as a process that is responsible for GEs \citet[e.g.][]{2001MNRAS.325..584S,2010arXiv1011.1222K,2013MNRAS.436.2484K}.
In \cite{2025ApJ...986..188M}, we studied a grid of massive accreting stars ranging from $\rm 20$ to $\rm 60~M_{\odot}$ with accretion rates varying from  $\rm 10^{-4}$ to $\rm 10^{-1} ~M_{\odot}~\rm yr^{-1}$ over a period of $20$ $\rm years$. We investigated how massive stars within the given range respond to accreting material, and how the accretion process alters their internal structure and evolutionary path. Using the \textsc{mesa} stellar evolution code, the star was modeled as a single star. We found that for high accretion rates $\rm \gtrsim 0.01 ~M_{\odot}~\rm yr^{-1}$, the companion star experienced a sudden increase in luminosity by about one order of magnitude, inflated, and moved towards the cooler side of the HR diagram. In contrast, for smaller accretion rates, the star remained on the hotter side of the HR diagram with a much smaller increase in luminosity during the accretion phase. It was based on the mechanics that, during the observation of GEs the increased luminosity by an order of magnitude, caused by the accretion onto the companion star. The gravitational energy released by this accretion powered the GEs and resulted in a high luminosity \citep[e.g.,][]{2001MNRAS.325..584S, 2010ApJ...723..602K, 2016RAA....16..117S}.

In this paper, we continue to develop the idea that massive stars can accrete mass at very high rates, and explore the effect of metallicity on the evolution of massive stars during the accretion phase.
The primary goal is to examine the companion's behavior in response to the mass accretion process, with calculations carried out within the framework of single stellar evolution models.
We deal only with post-MS stars and with accretion rates ($\dot{M}_{\rm acc}$) of up to $\rm \approx 0.01~M_{\odot}~yr^{-1}$. We propose the idea of accretion onto the star's outer layers to study at which threshold value of accretion disequilibrium occurs, i.e., the star switches from blue to redder. Simultaneously, we analyze the stellar properties at specific points of accretion. We also check the metallicity effect on the accretion luminosity and run simulations for both stars at Galactic, LMC, and SMC metallicities, with accretion rates ranging from  $\rm 10^{-5}$ to $\rm 10^{-2}~M_{\odot}~yr^{-1}$.

The paper is organized as follows.
Section \ref{2} discusses the basic assumptions and modeling method. Section \ref{3} presents the results of our analysis. Finally, Section \ref{4} provides our conclusions and a summary of the findings.

\section{Physical Ingredient of the models}
\label{2}
\begin{figure}
    \centering
\includegraphics[trim={0.15cm 0.0cm 0.8cm 0.0cm},clip, width=1\columnwidth]{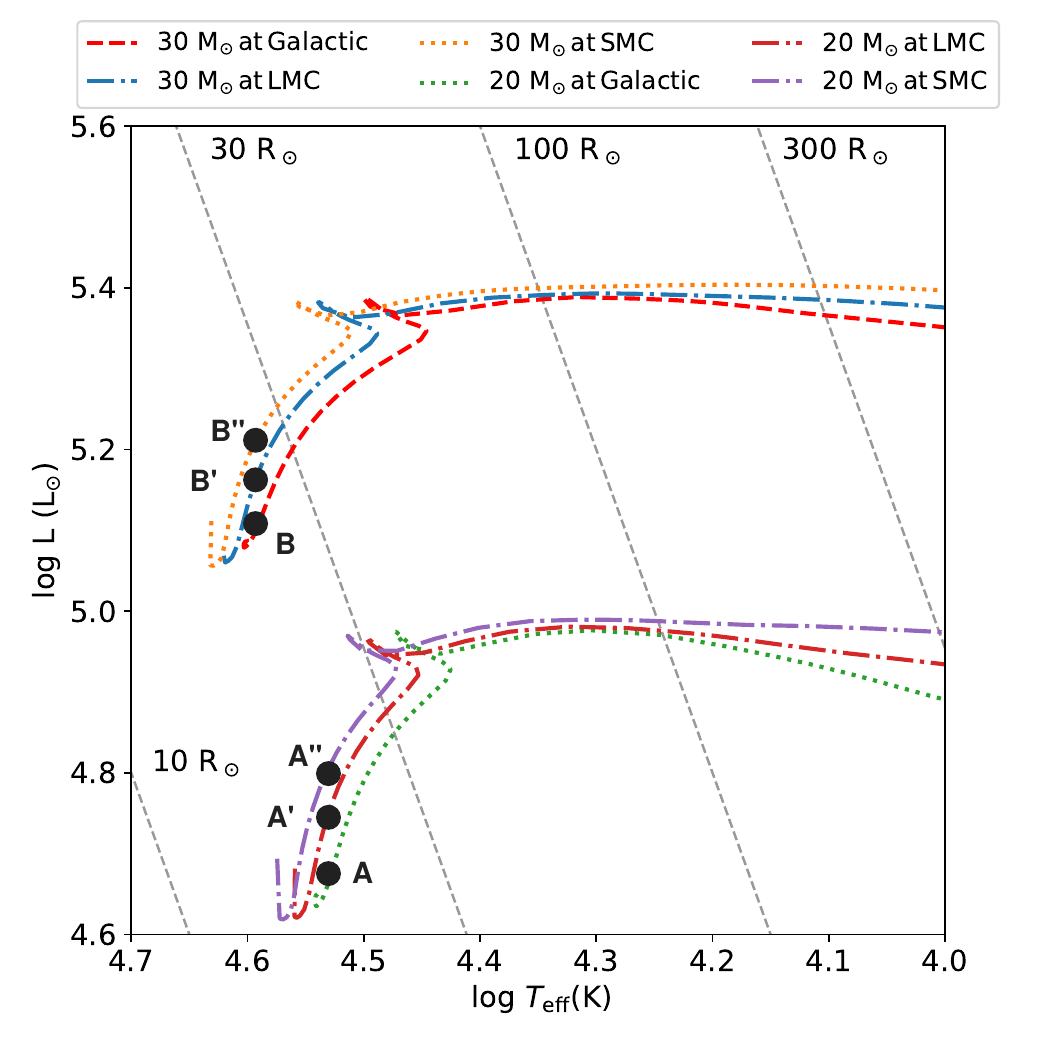}
     \caption{The evolutionary tracks of 20 and 30 $\rm M_{\odot}$ stars at Galactic, LMC, and SMC metallicities, spanning from the ZAMS to the helium-core burning phase. Points A, A', and A'' mark the initiation of accretion for the 20 $\rm M_{\odot}$, while points B, B', and B'' represent the initiation of accretion for the 30 $\rm M_{\odot}$ star.}
    
    \label{f1}
 \end{figure}

\subsection{Inital set up}
We simulate two models: in the first model, we evolve a $\rm 20~M_{\odot}$ non-rotating star along with Galactic, LMC, and SMC metallicities. In the second model, we evolve a $\rm 30~M_{\odot}$ non-rotating star along with Galactic, LMC, and SMC metallicities.  We assume that both of these stars are accreting companion stars in a binary system with a more massive star being the primary. We further assume that the primary undergoes a GE, during which mass is being lost from its surface at a high rate. The close orbit is assumed to result in accretion at a high rate of primary material onto the companion during the event of an eruption \citep{2010ApJ...709L..11K}. Our simulation does not treat the primary but focuses on the accreting companions as a single star, and follows its evolution under the influence of accretion. Orbital effects such as modulation of accretion rate in eccentric orbits \citep{2010MNRAS.405.1924K} are not modeled, and we assume that the accretion rate is constant. The calculations are performed with the stellar evolution code Modules for Experiments in Stellar Astrophysics  (\textsc{MESA} version- r23.05.1) \citep[e.g.,][]{2011ApJS..192....3P, 2013ApJS..208....4P, 2015ApJS..220...15P,2018ApJS..234...34P,2019ApJS..243...10P}. Accretion is initiated at points A, A', and A'' for our $\rm 20~M_{\odot}$, and at points B, B', and B'' for our $30~\rm M_{\odot}$ star.

Points A, A', and A" mark the evolutionary stages of the $20 \rm M_{\odot}$ star at different metallicities: A corresponds to Galactic metallicity (Z=0.02), A' to LMC metallicity (Z=0.008), and A" to SMC metallicity (Z=0.004). These points highlight how the internal composition or metallicity significantly influences the star’s position in the HR diagram as shown in Figure \ref{1}, reflecting changes in the stellar parameters as well. Similarly, points B, B', and B" represent the corresponding evolutionary stages for the $30 \rm M_{\odot}$ star for all three metallicities.

Stellar parameters corresponding to these points are given in Tables  \ref{T1} and \ref{T2}.  
The stars begin to accrete mass at varying accretion rates ranging from $ 10^{-5} $ to $ 10^{-2}~\rm M_{\odot}~yr^{-1}$, and a total of 234 points of accretion rates are simulated for both stars at Galactic, LMC, and SMC. In every metallicity value, there are 39 accretion rates ranging between  $\rm 10^{-5}$ to $\rm 10^{-2}~M_{\odot}~yr^{-1}$. The mass accumulation occurs in the outer layers of the star on a timestep ($ \simeq 500~\rm s$) that is shorter than the dynamic timescale of the star ($  t_{\rm dyn} \simeq 5 \times 10^{3} ~\rm s$). Tables \ref{T3} and \ref{T4} show the variation in the stellar parameters caused by the accretion scenario.

\subsection{MESA Modeling}

We start to evolve the $20$ and $30~\rm M_{\odot}$  star from the zero-age main sequence (ZAMS). During the evolution of the companion star from the pre-MS phase, we consider the presence of strong stellar winds by using the \textsc{mesa} `Dutch’ wind prescription with a scaling factor of 0.5 \citep{2020MNRAS.495..249Z}. During the accretion scenario, we do not consider the mass loss by stellar wind.  In \textsc{mesa}, the parameter for the initial mass fraction is defined as $ Y = Y_{\rm prim} + (\Delta  Y /\Delta Z) / Z$, where $ Y_{\rm prim}$ represents the He abundance,  $ Z$ is the metallicity and $ \Delta Y$ and $ \Delta Z$ represent the changes in them during the evolution. The values we use in our model are $ Y_{\rm prim}=0.24$ and $(\Delta  Y / \Delta  Z) / Z =2$, consistent with the default values in \textsc{mesa} \citep{2009yCat..72980525P}.
To model convection mixing, we employ the Mixing Length Theory (MLT) proposed by \citet{1965ApJ...142..841H}, with mixing length parameter, $\alpha_{\rm MLT}$  = 1.5. Semi-convection is modeled with an efficiency parameter $\rm \alpha_{sc}= 0.01$ \citep{1985A&A...145..179L}. Following \citet{1980A&A....91..175K}, we select the efficiency parameter for the thermohaline mixing to be $\rm \alpha_{th} = 2.0$.
We use \citet{2000A&A...360..952H} diffusive technique to represent the convective overshooting, using $f_{1}$ = 0.005 and $f_{0}$ = 0.001 for each convective core and shell. We use an opacity table of type II \textsc{OPAL} which allows for time-dependent variation in the metal abundance in our model \citep{1996ApJ...464..943I}. Massive stars exhibit convective cores and radiative envelopes. The presence of a radiative envelope leads to enhanced thermal energy within the envelope when mass accumulates on the star's surface, consequently causing the star to expand and deviate from its expected evolutionary trajectory. Simulating accretion in \textsc{mesa} requires a special approach that handles the interaction of the arriving material with the outer layers of the star. In each step, gravitational energy is added to the gas in the outermost cell. The method of how accretion is treated onto the outer layer of the star follows \citet[e.g.,][]{2011ApJS..192....3P, 2013ApJS..208....4P}. During the accretion phase the luminosity is given by \citep{2024ApJ...966L...7L}: 

\begin{equation}
     L_{\rm acc} = \frac{Gm{\dot{M}_{\rm acc}}}{R},
     \label{eqL}
\end{equation}

Here m is the mass of the star, R is the stellar radius, and $\dot M_{\rm acc}$ is the accretion rate to infall the material onto the outer layers of the star. Since angular momentum can impact the accretor star in terms of mass transfer efficiency, which might lead to significant changes in stellar structure \citep[e.g.,][]{2005A&A...435.1013P, 2022A&A...659A..98S}. It has been intentionally omitted from consideration in this study. Similarly, our model omitted stellar rotation. It is well-known that mass transfer in binary systems causes the companion star to spin up to near-critical rotation \citep{2013ApJ...764..166D}. \citet{1981A&A...102...17P} demonstrated that only a small amount of matter is needed to spin the star up to critical rotation. Consequently, all models should reach critical rotation very quickly. It is believed that induced rotational mixing \citep[e.g.,][]{2000ApJ...528..368H, 2005ApJ...626..350H} may help incorporate the incoming material into the star, potentially preventing or enhancing the expansion of the companion. Thus, there is uncertainty in our models for the accretion-induced expansion, and the amount of accumulated material on the surface depends on the mass accretion rate and period only. 

In these simulations, it is also assumed that the accreted material on the star's surface possesses the same composition as the material already present on the surface. Additionally, it is considered that the entropy of the accreting material aligns with the surface entropy of the accretor star \citep[e.g.,][]{2021ApJ...923..277R, 2023A&A...669A..45T}. \citet{2015ApJS..220...15P} gives a timescale argument which indicates that this is always true. However, according to the stellar structure equations \citep{2013sse..book.....K}, a change in entropy affects the luminosity and the radius of the star. Whether the expansion is enhanced or diminished depends on the actual entropy of the material, which is controlled by the initial entropy and the accretion mechanism.

\begin{table}
    \centering
    \begin{tabular}{l c c c c c c c c c}
    \hline 
    \hline
     Metallicity & Galactic & LMC & SMC\\
     \hline
      $ \rm star~age~ (Myrs) $ & 2.666112 & 5.695176 & 6.269452 \\
      $ M ( \rm M_{\odot})$ & 19.95  & 19.93 & 19.94 \\ 
     $\log T_{\rm eff}~ (\rm K)$ &  4.52 & 4.52 & 4.52 \\
     $\log R ~(\rm R_{\odot}) $ &  0.84 & 0.89 & 0.89 \\
     $\log g ~(\rm cm~s^{-2})$ & 4.06 & 3.94 & 3.96 \\
     $\log \rho ~(\rm g~cm^{-3})$ & -9.09 &  -9.14 & -9.13\\
     $\log L ~(\rm L_{\odot})$ & 4.71 & 4.80 & 4.82\\
     $\log \dot{M}~(\rm {M_{\odot}}\rm yr^{-1})$ &  -7.67 &  -7.76 & -7.90\\
    \hline 
     \hline
    \end{tabular}
    \caption{Stellar parameters for the $ \rm 20~M_{\odot}$ star at the initial stage of the accretion for all metallicities shown above. Here rows are from first to last: star age, the initial mass of the star ($ M$), surface temperature ($ T_{\rm eff}$), surface gravity ($ g $), density ($ \rho$), surface luminosity ($ L$), and mass loss via stellar winds ($ \dot{M}$) respectively. Here column 1 shows the stellar parameters, column 2, column 3, and column 4 show the stellar properties corresponding to the points A, A', and A''in Fig. \ref{f1} respectively. }
    \label{T1}
\end{table}

\begin{table}
    \centering
    \begin{tabular}{l c c c c c c c c c}
    \hline 
    \hline
     Metallicity & Galactic & LMC & SMC \\
     \hline
     $ \rm star~age~ (Myrs) $ & 1.1594326 & 2.8673181 & 3.869637\\
     $  M ~( \rm M_{\odot})$ & 28.85  & 29.79 & 29.80 \\ 
     $\log T_{\rm eff}~ (\rm K)$ &  4.58 & 4.58 & 4.58\\
     $\log R ~(\rm R_{\odot}) $ &  0.910 & 0.93 & 0.96\\
     $\log g ~(\rm cm~s^{-2})$ & 4.09 & 4.03 & 3.98\\
     $\log \rho ~(\rm g~cm^{-3})$ & -9.17 & -9.18 & -9.4\\
     $\log L ~(\rm L_{\odot})$ & 5.12 & 5.18 & 5.23\\
     $\log \dot{M}~(\rm {M_{\odot}}\rm yr^{-1})$ &  -6.87 & -7.02 & -7.12 \\
    \hline 
   \hline
    \end{tabular}
    \caption{Same as Table \ref{T1}, but for the $ \rm 30~M_{\odot}$ star.}
    \label{T2}
\end{table}

\begin{figure*}
  \centering
  \begin{tabular}{c @{\qquad} c }
    \includegraphics[width=.5\linewidth]{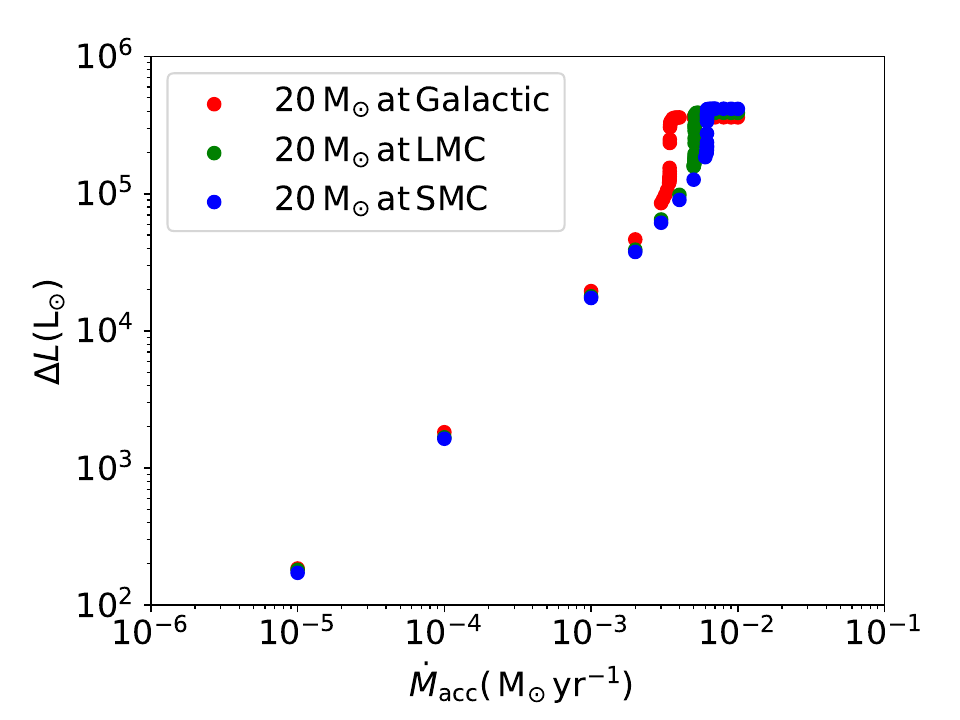}
    &
     \hspace{-.8cm} 
    \includegraphics[width=.5\linewidth]{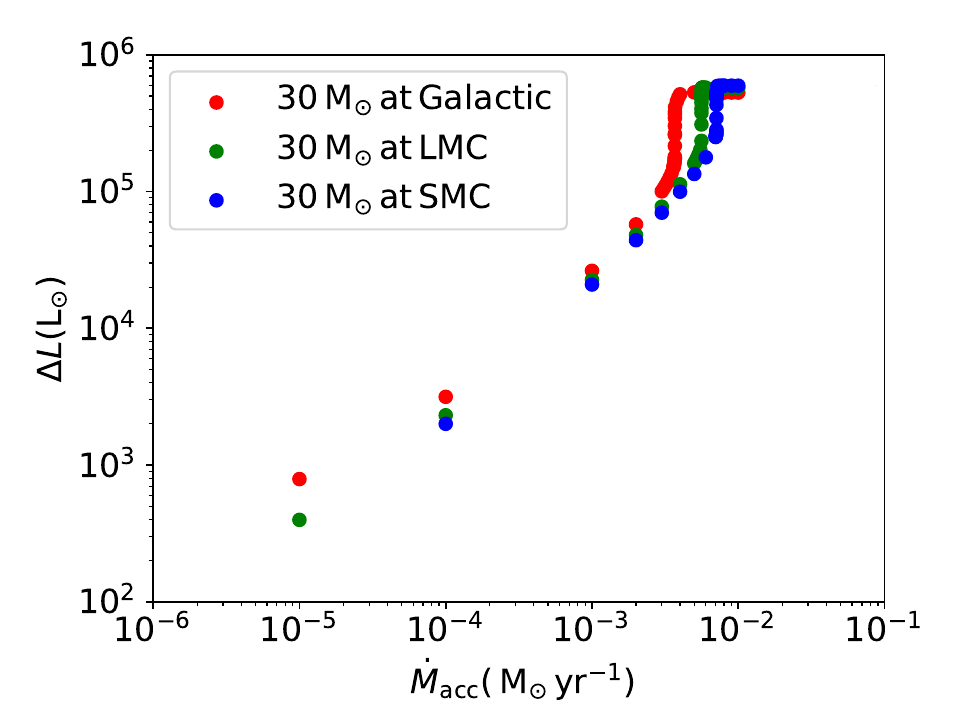}
    \\
    \small (a) & (b)
  \end{tabular}

\caption{Panels (a) and (b) show the increased luminosity ($\Delta L$) for a $20$ and $\rm 30~M_{\odot}$ star during its final phase of accretion, with varying accretion rates respectively. Here, red points correspond to the Galactic, green points correspond to the LMC, and the blue points correspond to the SMC metallicity, respectively.}
 \label{f2}
\end{figure*}

\section{Results}
\label{3}
We run two models: model 1 for  $\rm 20~M_{\odot}$ at Galactic, LMC  and SMC metallicities, and model 2 for $\rm 30~M_{\odot}$ stars at Galactic, LMC \& SMC metallicities with varying accretion rates. Fig. \ref{f2} presents an overview of the complete accretion simulations for all 239 runs.  Here Fig. \ref{f2} panel (a) shows the increased luminosity ($\Delta L$) for  $\rm 20~M_{\odot}$  at Galactic, LMC \& SMC, during the accretion phase along with the $\dot M_{\rm acc}$. Panel (b) shows the increased luminosity ($\Delta L$) for  $\rm 30~M_{\odot}$  at Galactic, LMC \& SMC during the accretion phase.

\begin{figure*}
  \includegraphics[trim= 0.0cm 0.0cm 0cm 0cm,clip=true,width=1\textwidth]{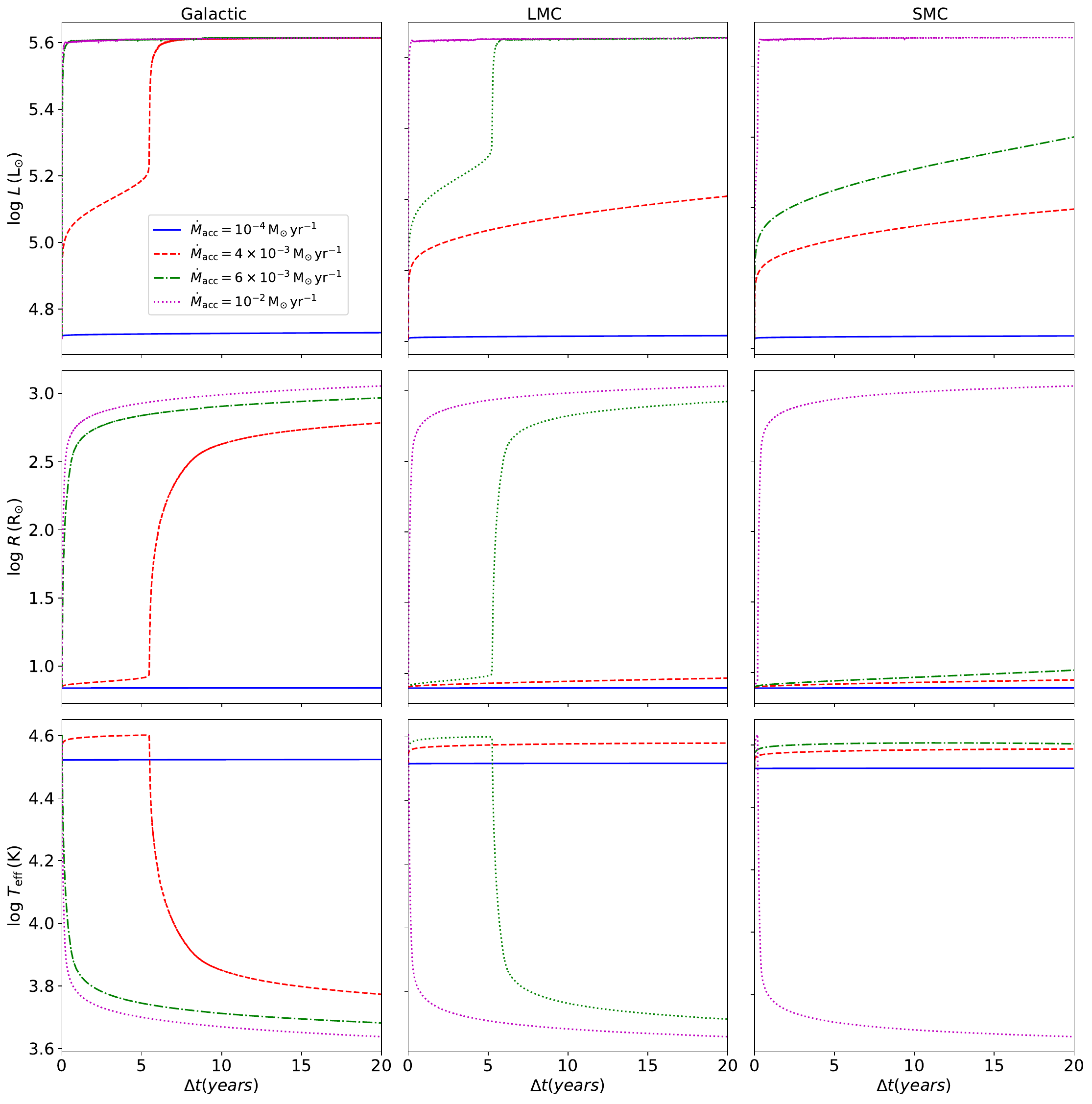} 
  \centering
\caption{The variation in luminosity $L$ (row 1), radii $R$ (row 2), and 
temperature $T_{\rm eff}$ (row 3) of the $\rm 20~M_{\odot}$ star at Galactic (column 1), LMC (column 2) and SMC (column 3) metallicities during the mass accretion process, respectively. At Galactic for $ \dot M_{\rm acc}= \rm 10^{-4}~M_{\odot}~yr^{-1} $ the star remains on the blue side of the H-R diagram and becomes hotter, while for $\rm 4 \times 10^{-3}$, $6 \times 10^{-3}$, and $\rm 10^{-2}~M_{\odot}~yr^{-1}$ accretion rates star becomes a cooler and inflate. At LMC for accretion rate  $\rm  10^{-4}$, and $ \rm 4 \times 10^{-3}~M_{\odot}~yr^{-1} $ the star remains on the blue side of the H-R diagram and becomes hotter, while for $6 \times 10^{-3}$, and $\rm 10^{-2}~M_{\odot}~yr^{-1}$ accretion rates star becomes a cooler and inflate. At SMC for accretion rates $\rm  10^{-4}~M_{\odot}~yr^{-1} $,  $ \rm 4 \times 10^{-3}~M_{\odot}~yr^{-1}$, and  $ \rm 6 \times 10^{-3}~M_{\odot}~yr^{-1} $ the star remains on the blue side of the H-R diagram and becomes hotter, while for $\rm 10^{-2}~M_{\odot}~yr^{-1}$ accretion rate star becomes a cooler and inflate.}
\label{f3}
\end{figure*}

\begin{figure*}
  \includegraphics[trim= 0.0cm 0.0cm 0cm 0cm,clip=true,width=1\textwidth]{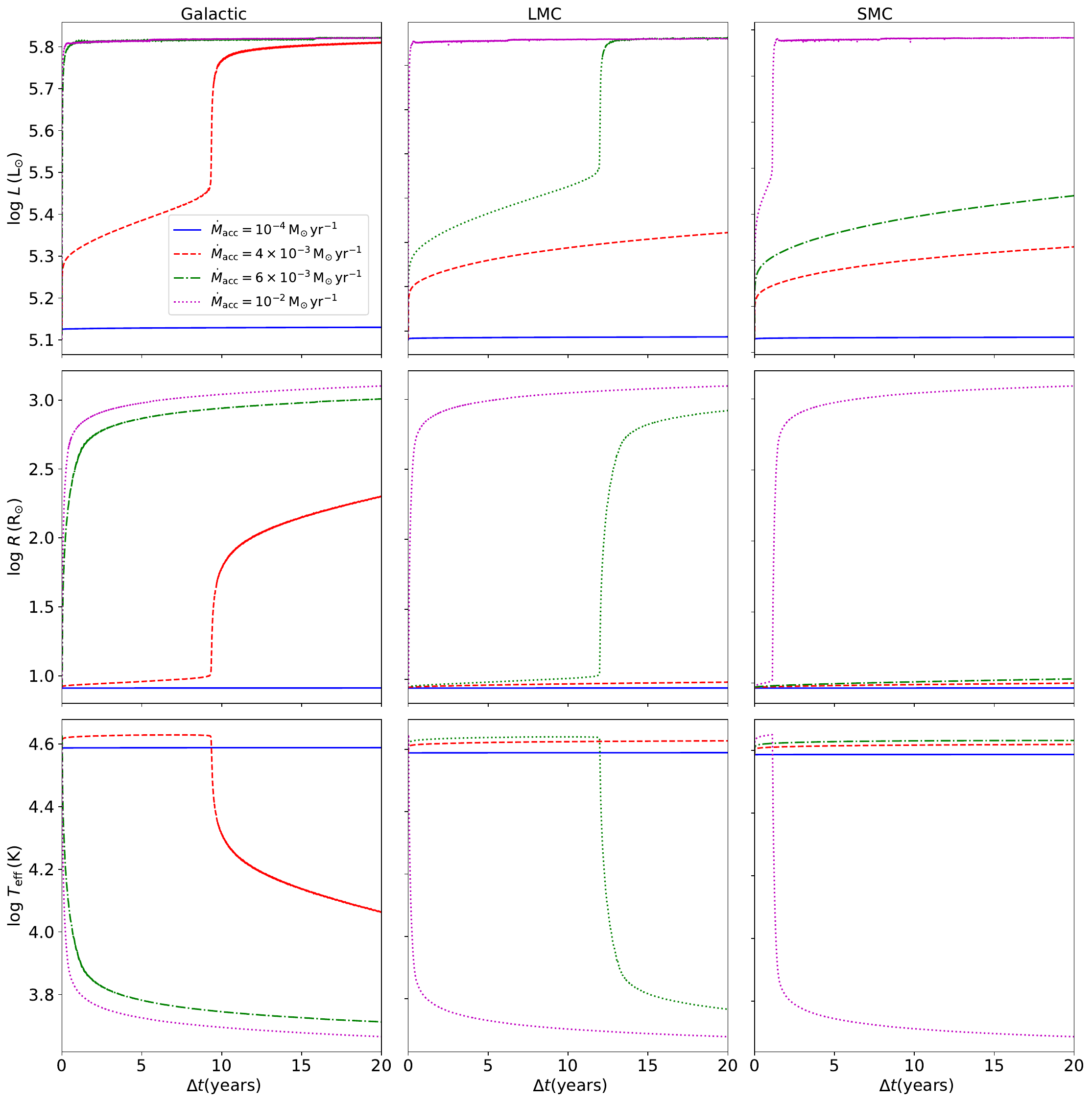} 
  \centering
\caption{Same as Figure \ref{f3}, but for the $ \rm 30~M_{\odot}$ star.}
\label{f4}
\end{figure*}

\subsection{20 and $ 30~\rm M_{\odot}$ companions response to the accretion }
We perform a simulation of 117 accretion rates, for a $\rm 20~M_{\odot}$ star at Galactic, LMC $\&$ SMC, spanning a range of accretion rates from $\rm 10^{-5}$ to $\rm 10^{-2}~M_{\odot}~yr^{-1}$. At the initial stage of accretion, i.e., point A in Fig. \ref{f1}, the luminosity of a $\rm 20~M_{\odot}$ at Galactic metallicity is log $L (~\rm L_{\odot})$ = 4.71, as mentioned in Table \ref{T1}. As a result of accretion during the period of 20 years, the luminosity increases. Interestingly, as shown in Fig. \ref{f3} and values in Table \ref{T3}, the star with lower accretion rates follows smaller variations in luminosity and radius, remaining on the hot side of the HR diagram during the accretion phase. In contrast, higher accretion rates lead to a significant increase in luminosity and radii values, resulting in a red and inflated star. We observe four distinct jumps in luminosity value: the first occurs at accretion rates between  $\rm 10^{-3}$ and  $\rm  10^{-2}~M_{\odot}~yr^{-1}$,  where the star shifts from hotter to cooler side of the HR diagram. Followed by a second jump occurs between $\rm 3 \times 10^{-3}$ to $\rm 4 \times 10^{-3}$, and a third jump occurs between $\rm 3.4 \times 10^{-3}$ to $\rm 3.5 \times 10^{-3}$, and fourth jumps occurs between $\rm 3.43 \times 10^{-3}$ to $\rm 3.44\times 10^{-3}~M_{\odot}~yr^{-1}$.  As mentioned in \citet{2025ApJ...986..188M}, the star maintains its thermal equilibrium for lower accretion rates, and there is a very small variation in the luminosity. They remain on the hot side of the HR diagram. Whereas for higher accretion rates, the star is driven out of thermal equilibrium in the initial days of accretion, where its luminosity increases by an order of magnitude. Later on, they inflate and become cooler stars. The luminosity and radii values corresponding to these jumps are mentioned in Table \ref{T3}.

Similarly at the LMC metallicity, for the  $\rm 20~M_{\odot}$, at point A' in Fig. \ref{f1}, we start the accretion. The luminosity of the star at this point is log $L ~(\rm L_{\odot})$ = 4.80. The star's luminosity increases as a result of the accretion of material during the 20 years. The star's behavior is similar to that of Galactic metallicity, with the lower accretion rate star remaining on the hotter side of the HR diagram. In contrast, the higher accretion rate star becomes cooler and inflated. For the LMC metallicity, the disequilibrium in stars' behavior occurs at different accretion rates. For the LMC simulations, these jumps occur: the first occurs at accretion rates between  $\rm  10^{-3}$ to  $\rm 10^{-2}$, followed by a second jump between $\rm 5 \times 10^{-3}$ to $\rm 6 \times 10^{-3}$, third jump occurs between $\rm 5.1 \times 10^{-3}$ to  $\rm 5.2 \times 10^{-3}$, and fourth jump occurs between $\rm 5.07 \times 10^{-3}$ to  $\rm 5.08\times 10^{-3}~M_{\odot}~yr^{-1}$ as mentioned in Table \ref{T3}. Meanwhile, for the SMC metallicity, we begin the accretion at point A'' in Fig. \ref{f1}, where the star's luminosity is log $L(\rm L_{\odot})$ = 4.83. Over 20 years, the star's luminosity increases due to the accretion of material. The star's behavior is similar to that of Galactic and LMC metallicities. The star with a lower accretion rate remains on the hotter side of the HR diagram, while a star with a higher accretion rate becomes cooler and more inflated. For the SMC, the disequilibrium occurs at different accretion rates. The first jump occurs between $\rm 10^{-3}$ and $\rm 10^{-2}$, the second jump occurs between $\rm 6 \times 10^{-3}$ and $\rm 7 \times 10^{-3}$, the third jump occurs between $\rm 6.1 \times 10^{-3}$ and $\rm 6.2 \times 10^{-3}$, and the fourth jump occurs between $\rm 6.16 \times 10^{-3}$ and $\rm 6.17\times 10^{-3}~M_{\odot}~yr^{-1}$, as detailed in Table \ref{T3}.

 Model 2 simulates the evolution of a $\rm 30~M_{\odot}$ at Galactic, LMC $\&$ SMC metallicities with varying accretion rates, ranging from $\rm 10^{-5}$ to $\rm 10^{-2}~M_{\odot}~yr^{-1}$. The accretion process starts at point B in Fig. \ref{f1} for Galactic metallicity value, where the luminosity of the star is log $L (\rm L_{\odot})$ = 5.12 as given in Table \ref{T2}. The luminosity of the $\rm 30~M_{\odot}$ star at point B is higher compared to the luminosity of the $\rm 20~M_{\odot}$ star at point A, due to the mass difference between the two stars. Here also, the star exhibits a similar behavior to $\rm 20~M_{\odot}$ star.  Although the jump occurs at different accretion rates compared to $\rm 20~M_{\odot}$ stars. The first jump occurs at $\dot M_{\rm acc}$ between  $\rm  10^{-3}$ to $\rm  10^{-2 }$, the second jump occurs at $\dot M_{\rm acc}$ between  $\rm 3 \times  10^{-3}$ to $\rm  4 \times 10^{-3}$, third jump occurs at $\dot M_{\rm acc}$ between  $\rm 3.6 \times 10^{-3}$ to $\rm 3.7 \times 10^{-3}$, and forth jump occurs at $\dot M_{\rm acc}$ between  $\rm 3.68 \times 10^{-3}$ to $\rm 3.69 \times 10^{-3}~M_{\odot}~yr^{-1}$.

Similarly, we simulate the evolution of  $\rm 30~M_{\odot}$ at LMC and SMC metallicities, during the accretion phase, where the accretion rate varies from $\rm 10^{-5}$ to $\rm 10^{-2}~M_{\odot}~yr^{-1}$. The luminosity of the star at points B' and B'' (see \ref{f1}) is log $L (\rm L_{\odot})$ = 5.18, and 5.23 as mentioned in Table \ref{T2}. The star shows similar behavior to Galactic for both metallicities. The mechanism for this is already mentioned in the section above. Although, the jump occurs at different accretion rates compared to the jumps at Galactic. For the LMC, a star switches from a hotter to a cooler star. We evolved, the first jump occurs at $\dot M_{\rm acc}$ between  $\rm  10^{-3}$ to $\rm  10^{-2 }$, the second jump occurs at $\dot M_{\rm acc}$ between  $\rm 5 \times  10^{-3}$ to $\rm  6 \times 10^{-3}$, third jump occurs at $\dot M_{\rm acc}$ between  $\rm 5.6 \times 10^{-3}$ to $\rm 5.7 \times 10^{-3}$, and forth jump occurs at $\dot M_{\rm acc}$ between  $\rm 5.68 \times 10^{-3}$ to $\rm 5.69 \times 10^{-3}~M_{\odot}~yr^{-1}$. While for the SMC, the first jump occurs at $\dot M_{\rm acc}$ between  $\rm  10^{-3}$ to $\rm  10^{-2 }$, the second jump occurs at $\dot M_{\rm acc}$ between  $\rm 7 \times  10^{-3}$ to $\rm  8 \times 10^{-3}$, third jump occurs at $\dot M_{\rm acc}$ between  $\rm 7 \times 10^{-3}$ to $\rm 7.1 \times 10^{-3}$, and forth jump occurs at $\dot M_{\rm acc}$ between  $\rm 7.09 \times 10^{-3}$ to $\rm 7.10 \times 10^{-3}~M_{\odot}~yr^{-1}$.

\subsection{The effect of the metallicity on the accretion luminosity}
We compare how the star in the LMC and SMC exhibits a distinct behavior in its increased accretion luminosity, as shown in Figure \ref{f2}, compared with the Galactic star. The LMC and SMC stars' luminosity is higher at the initial stage of the accretion phase as shown in Fig. \ref{f1}, compared to the Galactic due to lower metallicity. While the Galactic star's luminosity changes (log $\Delta L$) more significantly during the entire accretion period (see Table \ref{T3}).
We know that the difference in luminosity occurs due to the opacity behavior of the star's outer layers during the accretion phase. The higher metallicity in stars leads to increased opacity, which affects the efficiency of radiation transfer in the stellar interior. Thus, the star with higher metallicity has more elements in its atmosphere that can absorb and radiate energy, leading to a stronger line blanketing effect. This causes the stellar atmosphere to expand more. As follows equation \ref{eqL}, the change in luminosity depends on $R$ value, as other parameters are constant during the simulation.

 According to Eq. \ref{eqL}, the accretion luminosity $\Delta L_{\rm acc}$ is inversely proportional to the stellar radius $R$. So, if the star's envelope expands more, as it does in the Galactic case due to higher metallicity, we would expect $ \Delta L_{\rm acc}$ to be lower, since the radius is larger. However, Tables \ref{T3} and \ref{T4} show the opposite: the Galactic star experiences a larger increase in luminosity during the accretion phase compared to the LMC and SMC stars. This might seem counterintuitive at first, but the reason lies in how metallicity affects opacity. Higher metallicity increases the opacity in the outer layers of the star. This means that energy released by the accreting material is trapped more efficiently and takes longer to escape. As a result, the energy builds up and is released over time as additional luminosity. Even though the star with Galactic metallically has a larger radius, the higher opacity causes more energy to be retained and radiated, leading to a higher $\Delta L_{\rm acc}$. Therefore, the observed increase in luminosity at higher metallicity is not simply due to changes in radius, but rather due to the stronger effect of opacity in trapping and releasing energy during the accretion phase. This explains why both $R$ and $L$ increase with metallicity, and why the Galactic star shows the highest $\Delta L_{\rm acc}$ despite having the most extended envelope.

\subsection{The evolution of surface properties during the accretion phase}

We select 4  specific accretion rates from the luminosity track depicted in Fig. \ref{f2} and conduct a detailed study of stellar parameters during the accretion process. We also investigate the effects of metallicity on the envelope properties of massive stars. These accretion rates are: $\rm 10^{-4}$, $4 \times 10^{-3}$, $6 \times 10^{-3}$, and $\rm 10^{-2}~M_{\odot}~yr^{-1}$. Fig. \ref{f3}, row 1, row 2, and row 3 depict the variation in stellar parameters such as luminosity, radii, and temperature, respectively, for $\rm 20~M_{\odot}$  at Galactic (column 1), LMC (column 2), SMC (column 3) during the accretion phase. 
As we can see in Fig. \ref{f3}, the change in luminosity log$\Delta L$ during the accretion phase is indeed higher for Galactic compared to the LMC and SMC, as we mentioned in Table \ref{T3}, and the variation in the radius is also higher at Galactic. The luminosity changes for the Galactic by 0.001, 0.904, 0.906, and 0.904, while the radius changes by 0.008, 1.943, 2.126, and 2.214. For the LMC, the luminosity changes by 0.0003, 0.409, 0.856, and 0.855, and the radius changes by 0.0007, 0.076, 2.033, and 2.14. For the SMC, the changes in luminosity are 0.0001,0.370,0.575,0.863, and the changes in radii are 0.0001,0.057,0.126 and 2.145 for the accretion rates $\rm 10^{-4}$, $4 \times 10^{-3}$, $6 \times 10^{-3}$, and $\rm 10^{-2}~M_{\odot}~yr^{-1}$ respectively. The luminosity, radii, and temperature variation in Fig. \ref{f3} column 1 show that the star at Galactic becomes a cooler star at the accretion rate $\rm 4 \times 10^{-3}$,$\rm 6 \times 10^{-3}$, and $\rm  10^{-2}~M_{\odot}~yr^{-1}$, remains hotter for the accretion rate $\rm  10^{-4}~M_{\odot}~yr^{-1}$. While column 2 shows the star at LMC metallicity, shows a bit different behavior, for the accretion rates $\rm 6 \times 10^{-3}$, and $\rm  10^{-2}~M_{\odot}~yr^{-1}$, it becomes cooler and inflates while for the accretion rates $\rm  10^{-4}$, and $\rm 4 \times 10^{-3}~M_{\odot}~yr^{-1}$ it remains a hotter star. 
Column 3 shows that the star at SMC metallicity becomes cooler and inflates only for the accretion rates $\rm  10^{-2}~M_{\odot}~yr^{-1}$, and it remains a hotter star for the accretion rates  
$\rm 4 \times 10^{-3}$,$\rm 6 \times 10^{-3}$, and $\rm  10^{-2}~M_{\odot}~yr^{-1}$. Thus, metallicity plays a significant role, during the accretion process.

Fig. \ref{f4}, row 1, row 2, and row 3 depict the variation in stellar parameters such as luminosity, temperature, and radius, respectively, for $\rm 30~M_{\odot}$  at Galactic (column 1), LMC (column 2), and SMC (column 3) during the accretion phase.  For the Galactic, luminosity changes by 0.007, 0.690, 0.701, and 0.701, while the radius changes by 0.0008,1.391, 2.097, and 2.192, respectively. For the LMC, the luminosity changes by 0.001, 0.242, 0.681, and 0.681, and the radius changes by 0.0007, 0.043, 1.989, and 2.165.  For the SMC, the changes in luminosity are 0.0001,0.209,0.32, and 0.663 and the changes in radii are 0.0001,0.038,0.068 and 2.131 for the accretion rates $\rm 10^{-4}$, $4 \times 10^{-3}$, $6 \times 10^{-3}$, and $\rm 10^{-2}~M_{\odot}~yr^{-1}$ respectively.
Additionally, the luminosity, radii, and temperature variation in Fig. \ref{f4} column 1,  shows that the star at Galactic becomes a cooler star at the accretion rate $\rm 4 \times 10^{-3}$, $\rm 6 \times 10^{-3}$, and $\rm  10^{-2}~M_{\odot}~yr^{-1}$, remains hotter for the accretion rate $\rm  10^{-4}~M_{\odot}~yr^{-1}$. While column 2 shows the star at LMC metallicity, shows a bit different behavior, for the accretion rates $\rm 6 \times 10^{-3}$, and $\rm  10^{-2}~M_{\odot}~yr^{-1}$, it becomes cooler and inflates while for the accretion rates $\rm 10^{-4}$, and $\rm 4 \times 10^{-3}~M_{\odot}~yr^{-1}$ it remains a hotter star.
Column 3 shows that the star at SMC metallicity becomes cooler and inflates only for the accretion rates $\rm  10^{-2}~M_{\odot}~yr^{-1}$, and it remains a hotter star for the accretion rates  
$\rm 4 \times 10^{-3}$, $\rm 6 \times 10^{-3}$, and $\rm  10^{-2}~M_{\odot}~yr^{-1}$. This behavior is similar in both stars. Thus, this suggests that the star with higher metallicity undergoes the thermal disequilibrium at lower accretion rates compared to the lower metallicity stars. 

The initial luminosity of the star is higher at LMC and SMC metallicities compared to Galactic due to the composition, but the subsequent changes in luminosity during the accretion phase are more pronounced at Galactic. This suggests that the accretion luminosity not only depends on the radius of the star, as inferred from equation \ref{eqL}, but also on the metallicity of the star, as the changes in luminosity are unexpectedly higher at Galactic despite the larger radius change. If we see closely the value of $R$ and $L$ in Tables \ref{T3}, and \ref{T4}, the effect of the metallicity on the increased luminosity is more pronounced at the accretion  $ \gtrsim \rm 10^{-3}~M_{\odot}~yr^{-1}$.

Our models show that during accretion, the stellar luminosity increases significantly and approaches or even exceeds the Eddington luminosity. This occurs because the star is driven out of thermal equilibrium, leading to envelope expansion and a corresponding rise in radiative output. This behavior is consistent with our previous findings \citet{2025ApJ...986..188M}, where similar luminosity peaks were observed during the accretion phases.

\section{Summary and Discussion}
\label{4}

We simulate the response of a massive companion star in a binary system to the high-rate accretion of gas from a primary star undergoing a Giant Eruption. The mass accretion rate is influenced by the mass loss from the erupting star, the binary separation, and the radiation field of the accretor, which can reduce the amount of material being accreted \citep{1984ApJ...284..337O}. Similarly, the mass accretion rate could vary with changing orbital separation in an eccentric orbit. We adopt a simplified approach assuming a constant accretion rate. We then demonstrate how the star responds to the accumulation of mass in its outer layers, leading to changes in its surface stellar parameters.

In our simulations, we assume that the accreted material has the same chemical composition as the surface layers of the accreting companion star. In reality, the donor (primary) star may be in a more advanced evolutionary stage, potentially leading to differences in surface composition, particularly in nitrogen and other CNO-processed elements. This is especially relevant for massive binaries in close orbits, where the companion may accrete material enriched in nitrogen due to the evolution of the primary \citep{2019ApJ...870L..18Q}. However, in our study, both the primary (donor) and the companion (accretor) stars are assumed to be in the post-MS phase. At this stage, the surface of the donor star is still expected to be predominantly hydrogen-rich, with only modest enrichment of heavier elements such as nitrogen or carbon. Therefore, while assuming that the accreted material has the same composition as the companion's surface is a simplification, we consider it to be a reasonable approximation given the similar evolutionary stages of both stars.

\citet{2022MNRAS.516.3193K} conducted a systematic study of the accretion rate. They carried out a numerical experiment for ejecting winds in a massive colliding-wind binary system and quantified the accretion onto the secondary star under different primary mass-loss rates. They set a binary system comprising a luminous blue variable
(LBV) as the primary and a Wolf–Rayet (WR) star as the secondary, and vary the mass-loss rate of the LBV to obtain different values of the wind momentum ratio $\eta$. They analyzed conditions under which the secondary star, often the less massive one, can accrete material from the primary's wind. When the primary star’s wind is stronger, the secondary may accrete more of this material, depending on the wind momentum balance.
They also mentioned how changes in the wind momentum ratio influence the shape and position of the shock region, which directly affects accretion efficiency. If the wind momentum of the secondary is much weaker than the primary, the accretion rate increases.

We ran a numerical experiment to analyze the response of massive stars to the accretion process, and the effect of metallicity on the accretion luminosity. During the accretion process, the star gains a significant amount of material in its outer layers as a result its stellar structure changes. We used 1-D stellar evolution code \textsc{mesa} to study the accretion process during the post-MS phase of massive stars. We explicitly examined the effect of accretion rates on the stars and the transition rates at which a star switches from a hotter to a cooler state. 

In our previous study, \citet{2025ApJ...986..188M}, we presented a comprehensive grid of massive stars with initial masses ranging from 20 to 60 $\rm M_{\odot}$ undergoing accretion (see figure 8 in that work). That study focused on the structural response of stars to accretion but did not explore the role of metallicity or provide a detailed investigation of the critical accretion rate at which a star transitions from maintaining thermal equilibrium to expansion. For the current work, we selected the 20 $\rm M_{\odot}$ model as a representative case to study the effects of metallicity on the accretion response in detail. We later included the 30 $\rm M_{\odot}$ model to strengthen our results and verify the consistency of our findings across different stellar masses, as both star shows the same behavior. Presenting two models provides a broader context and enhances the generality of our conclusions, especially regarding the threshold accretion rates and their metallicity dependence.

Thus, we evolved 20 and $\rm 30~M_{\odot}$ stars from the ZAMS phase and introduced the accretion process during the post-MS phase. We ran two models in which a star switches from a hotter to a cooler star. We evolved: model 1 for  $\rm 20~M_{\odot}$  at Galactic, LMC $\&$ SMC metallicities, and model 2 for $\rm 30~M_{\odot}$ star at Galactic, LMC $\&$ SMC metallicities, with varying accretion rates. For every metallicity value, there are 39 simulations in which accretion rates ranged from $\rm 10^{-5}$ to $\rm 10^{-2}~M_{\odot}~yr^{-1}$.
\citet{2010ApJ...723..602K}, studied the companion star during the periastron passage triggering the 19$^{\rm th}$ century $\eta$ Carinae eruptions. They found that periastron passage led to a significant increase in the luminosity of the companion star by an order of magnitude for an accretion rate $\rm \simeq 0.1-0.2~M_{\odot}~yr^{-1}$. In our work, for the high accretion rates ($\rm  10^{-2}$), the increased luminosity is $\Delta \log L \simeq 0.70$ dex, which is the same order of magnitude. For the lower accretion, the change in $L$ is much smaller, as expected. 

We find that stars with lower accretion rates remain on the hot side of the HR diagram, as it remains in thermal equilibrium during the accretion scenario. While the star with higher accretion rates becomes cooler and inflate as a result of thermal disequilibrium, as mentioned in \citet{2025ApJ...986..188M}. The disequilibrium is also influenced by the mass of the star and metallicity, as our simulations show that the transition of the star from the hotter to a cooler side of the HR diagram occurs differently for each metallicity. 
As mentioned in Sec. \ref{3}, and Table \ref{T3} the jump where the star switches from hotter to cooler side, depends on the metallicity, as we observed that stars with higher metallicity become cooler stars with lower accretion rates compared to stars with lower metallicity, such as LMC, and SMC metallicities. 

\textit{Our analysis shows that the metallicity effects opacity and also influences the expansion of the companion star and the accreted luminosity. Higher metallicity increases opacity, which in turn decreases the Eddington accretion rate and increases the accretion luminosity.} We explore the discrepancies that arise when using Galactic and lower metallicities on to the accretion rates, focusing on their impact on the increased radius and luminosity of the companion star.
 Our analysis revealed that \textit{the variation in accretion luminosity is higher for higher-metallicity stars (Galactic) despite their larger radii, due to the increased opacity of their outer layers compared to those of lower-metallicity stars (LMC, and SMC)}. This suggests that during the accretion phase, the increased luminosity of the star is influenced by the opacity of its outer layers. 

\vspace{0.5cm}
We thank an anonymous referee for helpful comments. We acknowledge the Ariel HPC Center at Ariel University for providing computing resources that have contributed to the research results reported in this paper. We acknowledge support from the AGASS center in Ariel University. The \textsc{mesa} inlists and input files to reproduce our simulations and associated data products are available on Zenodo (10.5281/zenodo.15682592).

\section{Appendix}

Tables \ref{T3} and \ref{T4} provide the stellar parameters for 20 and 30 $\rm M_{\odot}$ stars at Galactic, LMC, and SMC metallicities, respectively. These parameters are derived from \textsc{mesa} profiles at the end of the accretion phase. The variations in these parameters show the impact of metallicity and accretion on the structural properties of the stars. 

\begin{sidewaystable*}
    \centering

   \begin{tabular}{l  c c c c| c c c c c |c c c c c } 
\hline
\hline
{} & \multicolumn{2}{c}{$\rm 20~M_{\odot}$ at Galactic} & \multicolumn{7}{c}{$\rm 20~M_{\odot}$ at LMC}  & & & $\rm 20~M_{\odot}$ at SMC \\
  \hline  
$\dot M_{\rm acc} $  & log $L$ &   log $\Delta L$ & log $R$  &log $\Delta R$ & $\dot M_{\rm acc} $& log $L$ &log $\Delta L$ & log $R$ &log $\Delta R$ & $\dot M_{\rm acc}$ & log $L$ &   log $\Delta L$ & log $R$  &log $\Delta R$ \\
\hline
\hline

    $ 10^{-2} $  & 4.715 &0.001& 0.840&0.008 &   $ 10^{-2} $  & 4.805 &0.0003 & 0.895 & 0.0007 &   $ 10^{-2} $   &4.825 &0.0001&0.888 & 0.0001 \\
    $ 10^{-1} $  & 4.728 &0.014 & 0.84 &0.01&   $ 10^{-1} $  & 4.815 &0.010 & 0.897 &0.001 &   $ 10^{-1} $ & 4.835& 0.010 &0.889 & 0.001 \\

    $ 1 $  & 4.856 &0.146 & 0.854 &0.014&   $ 1 $  & 4.911 &0.111 & 0.908&0.012 &  $ 1 $ & 4.924  &0.099 & 0.899  & 0.01\\

     2 & 4.991&0.281 & 0.875 &0.035 &   2   & 5.012 &0.212 & 0.923 &0.033 &  2  & 5.018 & 0.193& 0.912 & 0.024 \\
   3 & 5.137 &0.427& 0.910& 0.07& 3  &5.109 &0.309 &0.941  &0.051 &  3  & 5.107  & 0.282& 0.927& 0.038\\
    $ 3.1 $  &5.154&0.444 & 0.916 &0.076&  $ 4  $ & 5.209 &0.409 & 0.966 &0.076 & $ 4$ &5.195  &0.370  & 0.946 &   0.057\\
    $ 3.2  $  & 5.175 &0.465& 0.924 &0.084 &  $ 5  $ & 5.345 &0.545 & 1.021 &0.131 & $ 5 $ &5.286 &  0.461&0.971 &0.082   \\
    $ 3.3 $  & 5.198 &0.488& 0.934 &0.094&  $ 5.01 $ & 5.348 &0.548& 1.022&0.132 & $ 6 $ & 5.400 &0.575 & 1.015 & 0.126  \\
    $ 3.4  $  & 5.232 &0.522& 0.952 &0.112 &  $ 5.02  $ & 5.351 &0.551 & 1.024 & 0.134 & $ 6.1 $ &5.421 &0.596 & 1.026 &0.137 \\
   
    $ 3.41  $  & 5.239&0.529 & 0.956 &0.116  &  $ 5.03 $ & 5.354 &0.554 & 1.026 &0.136 & 6.11 & 5.423 & 0.598 &1.028 &0.139 \\
    $ 3.42  $  & 5.245 &0.535& 0.0.960 &0.12 &  $ 5.04  $ & 5.359 &0.559& 1.029&0.139 & 6.12 & 5.426 & 0.601 &1.03 &0.141  \\
    $ 3.43  $  & 5.255 &0.545& 0.966 &0.126 &  $ 5.05  $ & 5.363 &0.563& 1.032 & 0.142 & 6.13 &5.430 & 0.605 &1.032 &0.143  \\
    $ 3.431  $  & 5.256 &0.546& 0.967 &0.127 &  $ 5.06  $ &5.369&0.569 &1.037 &0.147 & 6.14 & 5.434 & 0.609 &1.035 &0.146 \\
    $ 3.432  $  & 5.260 &0.55& 0.970 &0.13  &$ 5.07  $ & 5.382&0.582 & 1.046 & 0.156   & 6.15 & 5.439 & 0.614 &1.038 &0.149 \\
     3.433   & 5.262 &0.552& 0.971 &0.131 &  5.071  & 5.384 & 0.584 & 1.048 & 0.158 & 6.16 &5.453 & 0.628 & 1.048 & 0.159  \\
    $ 3.434  $  & 5.263 &0.553& 0.973&0.133  & $ 5.072 $ & 5.390 & 0.59 & 1.053 & 0.163 &6.161 & 5.460 & 0.635 &1.055 &0.166 \\

    $ 3.435 $  & 5.266 &0.556& 0.975 &0.135 & $ 5.073 $ & 5.399 & 0.599 & 1.063 & 0.173 & 6.162 & 5.482 &0.657 &   1.081 &0.192 \\
    $ 3.436 $  & 5.272 &0.562 & 0.980 &0.14 & $ 5.074  $ &  5.413 & 0.613 & 1.078 & 0.188 & 6.163  & 5.533 & 0.708 &1.19 &0.301\\
    $ 3.437 $  & 5.281&0.571 & 0.987 &0.147 & $ 5.075 $ & 5.473 & 0.673 & 1.185 & 0.295   & 6.164 & 5.569 & 0.744 &1.327 &0.438  \\
    $ 3.438 $  & 5.313&0.603 & 01.023&0.183  & $ 5.076$ & 5.501 & 0.701 & 1.259 & 0.369 &6.165 & 5.605  &0.780 &1.469 &0.58  \\
    $ 3.439  $  & 5.456 &0.746 & 1.310 &0.47& $ 5.077 $  & 5.544 & 0.744 & 1.401 & 0.511 & 6.166 & 5.620 & 0.795 &1.557 &0.668\\
    $ 3.44  $  & 5.477&0.767 & 1.371 &0.531 &  $ 5.078 $ & 5.5628  &0.762 &  1.476 & 0.586  & 6.167  &5.634 & 0.809 &1.653 &0.764\\
    $ 3.45 $  & 5.552&0.842 & 1.684&0.844  & $ 5.079 $ & 5.573 & 0.773& 1.529 & 0.639 & 6.168 & 5.641 & 0.816 &  1.706 &0.817\\
    $ 3.46 $  & 5.567 &0.857 & 1.781 &0.941  & $ 5.08 $ & 5.582 & 0.782& 1.574& 0.684 & 6.169 & 5.648 & 0.823&1.771 &0.882\\
    $ 3.47 $  & 5.577 &0.867 & 1.855 &1.015 & $ 5.09  $ &5.629 &0.829& 1.962   & 1.072 & 6.17 & 5.655 &0.830 &1.831 &0.942\\
    $ 3.48  $  & 5.582&0.872 & 1.905 &1.065  &  $5.1  $ & 5.644 &0.844 & 2.204 &  1.314 & 6.18 & 5.678 & 0.853 &2.272 &1.383  \\
    $ 3.49  $  & 5.586 &0.876& 1.958&1.118  &$ 5.2  $ & 5.654 &0.854 & 2.716 & 1.826 & 6.19 & 5.680&0.855 &2.517 &1.628 \\
    $ 3.5  $  &  5.590 &0.88 & 2.003 &1.163  & $ 5.3 $ & 5.656 &0.856& 2.782 &1.892 & 6.2 &  5.681 & 0.861& 2.627 & 1.739 \\
    $ 3.6 $  &5.608 &0.898&2.373 &1.533  &$ 5.4  $ & 5.654&0.854 & 2.825 & 1.935 &6.3 & 5.682 &0.862 & 2.797 &1.909   \\
    $ 3.7  $  & 5.612 &0.902 &2.457 &1.617  & $5.5 $ & 5.653 & 0.853 & 2.851 & 1.961 &6.4 & 5.682 & 0.862& 2.848 & 1.959  \\
    $ 3.8 $  &5.613 &0.903 & 2.694 &1.854& $ 5.6$ & 5.653 &0.853 & 2.871 &1.981  & 6.5 & 5.682 &0.862 & 2.877 &1.989  \\
    $ 3.9  $  & 5.613 &0.903 & 2.746 &1.906 &$ 5.7 $ & 5.653 &0.853& 2.887 &1.997 & 6.6  & 5.682 &0.862 & 2.898 &2.01  \\
    $ 4  $  &5.614 & 0.904& 2.783&1.943 &$ 5.8  $ & 5.654&0.854 & 2.900 &2.01 & 6.7  & 5.683 &0.863 & 2.914 &   2.026\\
    $  5 $  & 5.615&0.905 & 2.920 &2.08 & $ 5.9 $ & 5.656 &0.8567 &2.914 &2.024  & 6.8  & 5.683 & 0.863& 2.929 &2.041  \\
    $  6  $  & 5.616 &0.906 &2.966 &2.1267 &$ 6  $ & 5.656&0.856 & 2.923 &2.033 & 6.9  &  5.683 &0.863 & 2.939 &2.051 \\
    $  7  $  & 5.615 &0.905 & 2.997 &2.157 &$ 7 $ & 5.655 &0.855& 2.977 &2.087 & 7 & 5.684 &0.864  & 2.948 & 2.06 \\
    $  8  $  & 5.616&0.906 & 3.020 &2.18& $ 8  $ & 5.655&0.855 & 3.002 &2.112 & 8 & 5.684 & 0.864 &2.997 & 2.108\\
    $  9 $  & 5.616&0.906 & 3.038 &2.198 & $ 9  $ & 5.655 &0.855& 3.033 & 2.143 & 9 & 5.683 &0.863 & 3.019 & 2.131  \\
    10  & 5.614 &0.904 & 3.054 & 2.214 & 10  & 5.655&0.855 & 3.033 & 2.14 & 10 & 5.683 &0.863 & 3.033 & 2.145\\

     \hline \hline
     \end{tabular}
    \caption{ The table presents the stellar parameters for the final profiles of the accretion phase over 20 years for a $\rm 20~M_{\odot}$ star at Galactic, LMC, and SMC metallicities. The columns include accretion rate ($\dot M_{\rm acc}$ in $\rm 10^{-3}~M_{\odot}~yr^{-1}$), luminosity (log $L$ in $\rm L_{\odot}$),  increased luminosity due to accretion (log $\Delta L$ in $\rm L_{\odot}$ ), radius (log $R$ in $\rm R_{\odot}$), and increased radius due to accretion (log $\Delta R$ in $\rm R_{\odot}$), respectively.}
    \label{T3}

   \end{sidewaystable*}

  \begin{sidewaystable*}
    \begin{tabular}{l c c c c| c c c c c| c c c c c}
\hline
\hline
{} &\multicolumn{2}{c}{$\rm 30~M_{\odot}$ at Galactic} & \multicolumn{7}{c}{$\rm 30~M_{\odot}$ at LMC}  & & $\rm 30~M_{\odot}$ at SMC\\

    \hline
    $\dot M_{\rm acc} $  & log $L$ & log $\Delta L$ & log $R$ &log $\Delta R$  & $\dot M_{\rm acc} $  & log $L$ &  log $\Delta L$ & log $R$ & log $\Delta R$   & $\dot M_{\rm acc} $  & log $L$ &  log $\Delta L$ & log $R$ & log $\Delta R$   \\

\hline
\hline

    $ 10^{-2} $  & 5.123 &0.007 & 0.910&0.0008 & $ 10^{-2} $  & 5.181 & 0.001& 0.936 &0.0007 &  $ 10^{-2} $ & 5.228 & 0.0001 & 0.963 & 0.0001 \\  
    $ 10^{-1} $  & 5.130  & 0.01& 0.911 & 0.001& $ 10^{-1} $  & 5.186  &0.006& 0.937 &0.0003 &  $ 10^{-1} $ & 5.233 & 0.005 & 0.964 & 0.0002\\
    $ 1 $  & 5.199  &0.079 & 0.921 & 0.011& $ 1 $  & 5.240 &0.06 & 0.944 &0.007 &  $ 1 $ & 5.278 & 0.058 & 0.970 & 0.0062\\

    $ 2 $  & 5.277  & 0.157& 0.936 & 0.026 &  $ 2  $  & 5.299 &0.119& 0.954 &0.024 &  $ 2  $ & 5.328 & 0.108 & 0.978 & 0.018 \\

    $ 3 $  & 5.365 & 0.245& 0.959 & 0.049&  $ 3  $  & 5.359 &0.179& 0.965 &0.035&  $ 3  $  & 5.378 & 0.158 & 0.987 & 0.027\\
     $ 3.1 $  &5.376 &0.256 & 0.962 &0.052 & $ 4$  & 5.422&0.242 & 0.979 &0.043 &  $ 4  $  & 5.429 & 0.209 & 0.998 & 0.038\\

    $ 3.2 $  & 5.387 &0.267& 0.966 & 0.056&   $ 5  $  & 5.494 &0.314& 1.00 & 0.07 &  $ 5  $  & 5.482 & 0.262 & 1.011 & 0.051\\
     $ 3.3 $  & 5.399 &0.279& 0.971 & 0.061& $ 5.1  $  & 5.503 &0.323& 1.003 &0.073 &  $ 6 $  & 5.540 & 0.32 & 1.028 & 0.068\\ 
  $ 3.4  $  & 5.412 &0.292& 0.977 & 0.067&  $ 5.2  $  & 5.512 &0.332& 1.007  & 0.077 &  $ 7 $  & 5.622 & 0.402 & 1.064 & 0.104\\
     $ 3.5  $  &  5.428 &0.308& 0.984 & 0.074& $ 5.3  $  & 5.522 &0.34& 1.011 &0.081 &  $ 7.01  $ & 5.624 & 0.404 & 1.065 & 0.105    \\
    $ 3.6 $  &5.448 & 0.328& 0.996 & 0.086 &  $ 5.4  $  & 5.534 &0.354& 1.017 &0.087 &  $ 7.02 $ & 5.625 & 0.405 & 1.066 & 0.106\\
    $ 3.61  $  & 5.451 &0.331& 0.998 &0.088& $ 5.5 $  & 5.548 &0.368& 1.024 &0.094  &  $ 7.03 $ & 5.627 & 0.407 & 1.067 & 0.107\\
    $ 3.62  $  & 5.454 & 0.334& 1.00 &0.09 & $ 5.6 $  & 5.586 &0.406& 1.050 &0.12 &  $ 7.04  $ & 5.629 & 0.409 & 1.068 & 0.108    \\
    $ 3.63  $  & 5.458 & 0.338&1.002 &0.092&   $ 5.601  $  & 5.663 & 0.483 & 1.175 & 0.245 &  $ 7.05 $  & 5.631 & 0.411 & 1.069 & 0.109\\
    $ 3.64  $  & 5.461 &0.341& 1.005 & 0.095&  $ 5.602  $  & 5.721 & 0.541 & 1.330 & 0.40 &  $ 7.06 $ & 5.634 & 0.414 & 1.071 & 0.111 \\
    $ 3.65 $  & 5.464 & 0.344& 1.007 & 0.097&  $ 5.603  $  & 5.744 & 0.564 & 1.405 & 0.475 &  $ 7.07$ & 5.637 & 0.417 & 1.073 & 0.113  \\
    $ 3.66  $  & 5.470 &0.35&1.011 & 0.101&  $ 5.604  $  & 5.776 &0.596 & 1.517  &0.587 &  $ 7.08 $ & 5.642 & 0.422 & 1.076 & 0.116   \\
     $ 3.67 $  & 5.476 &0.356 & 1.015 & 0.105 &  $ 5.605  $  & 5.797 & 0.617  &1.662 & 0.732 &  $ 7.09 $ & 5.650  & 0.43 & 1.083 & 0.123 \\
     $ 3.68  $  & 5.480 & 0.36 & 1.021 &0.111&  $ 5.606 $  &  5.804 & 0.624 & 1.708 & 0.778 &  $ 7.091 $ & 5.655 & 0.435 & 1.087 & 0.127   \\ 
      $ 3.681 $  & 5.484 &0.364 & 1.023 &0.113&  $ 5.607  $  & 5.809 & 0.629 & 1.748 & 0.818 &  $ 7.092 $ &  5.710 & 0.490 & 1.177 & 0.217 \\

      $ 3.682  $  & 5.488 &0.368 &1.026&0.116 &  $ 5.608 $  & 5.815 & 0.635 & 1.791 & 0.861 &  $ 7.093 $ &  5.777 & 0.557 & 1.386 & 0.426  \\
     $ 3.683 $  & 5.494 &0.374 & 1.032 &0.122 &  $ 5.609 $  & 5.820 & 0.64 & 1.836 & 0.906 &  $ 7.094 $ & 5.807 & 0.587 & 1.525 & 0.565  \\
     $ 3.684 $  & 5.541 & 0.421 & 1.089 &0.179 &  $ 5.61 $  & 5.824 & 0.644 & 1.871 & 0.941 &  $ 7.095 $ & 5.823 & 0.603 & 1.622 & 0.662 \\
     $ 3.685  $  & 5.592 & 0.472 & 1.182 &0.272&  $ 5.62$  & 5.842 & 0.662 & 2.140 & 1.21 &  $ 7.096 $ & 5.823 & 0.603 & 1.627 & 0.667 \\
     $ 3.686 $  & 5.597 &0.477 &1.193 & 0.283 & $ 5.63  $  & 5.851 & 0.671 & 2.335 & 1.405 &  $ 7.097$ &  5.829 & 0.609 & 1.672 & 0.712 \\

    $ 3.687  $  & 5.637 & 0.517&1.285 &0.375& $ 5.64  $  & 5.858 & 0.678 & 2.502 & 1.572 &  $ 7.098$ & 5.834 & 0.614 & 1.703 & 0.743  \\  
     $ 3.688$  & 5.678 & 0.558& 1.395 &0.485 & $ 5.65  $  & 5.860 & 0.68 & 2.596  & 1.666 &  $ 7.099$ &5.841 & 0.621 & 1.766 & 0.806  \\
    $ 3.689  $  & 5.698 & 0.578& 1.464 &0.554 & $ 5.66 $ & 5.860 & 0.68 & 2.658 & 1.728 & $ 7.1  $  &  5.846 & 0.626 & 1.813 & 0.853  \\
     $ 3.69  $  &5.711 & 0.591& 1.507 &0.597  & $ 5.67  $ & 5.860 & 0.68 & 2.698 & 1.768 & $ 7.2 $  & 5.880 & 0.662 & 2.839 & 1.879  \\
    $ 3.7  $  & 5.737 &0.617 &1.607 &0.697 & $ 5.68 $ &5.860 & 0.68 & 2.726 &1.796   & $ 7.3  $  & 5.881 & 0.661 & 2.905 & 1.945\\
    $ 3.8  $  &5.769 &0.649 & 1.799 & 0.889 & $ 5.69  $ & 5.859 & 0.679 &2.747 &1.817 & $ 7.4 $  & 5.881 & 0.661 & 2.939 &1.979 \\
    $ 3.9 $  & 5.793 &0.673 & 2.019 & 1.109 &  $ 5.7  $  & 5.860 &0.68& 2.764 &1.834 & $ 7.5  $  & 5.881 & 0.661 & 2.962 & 2.002\\
    $ 4  $  &5.810 &0.69 & 2.301 & 1.391& $ 5.8  $  & 5.860&0.68 & 2.850 &1.92 & $ 7.6  $  & 5.881 & 0.661 & 2.979 & 2.019\\
    $  5  $  & 5.820 &0.70 & 2.940 &2.03 & $ 5.9 $  & 5.861&0.681 & 2.892 &1.962 & $ 7.7 $  & 5.881 & 0.661 & 2.993 & 2.033 \\
    $  6  $  & 5.821 &0.701 &3.007 &2.097& $ 6 $  & 5.861&0.681 & 2.919 &1.989  & $ 7.8 $  & 5.881 & 0.661 & 3.00 & 2.04 \\
    $  7  $  & 5.821 &0.701  & 3.042 &2.132 &  $ 7  $  & 5.861&0.681 & 3.024 &2.094  & $ 7.9  $  & 5.881 & 0.661 & 3.013 & 2.053 \\
    $  8 $  & 5.820 &0.70& 3.066 &2.156 & $ 8  $  & 5.861&0.681 & 3.060 &2.13 & $ 8  $  &5.883 & 0.663 & 3.023 & 2.063\\ 
    $  9  $  & 5.821 &0.701 & 3.086 &22.176 &  $ 9  $  & 5.861&0.681 & 3.081 &2.151 &  $ 9  $  & 5.883 & 0.663 & 3.069 & 2.109 \\
    $ 10 $  & 5.821 &0.701 & 3.102 &2.192 & $ 10 $  & 5.861 &0.681& 3.095 &2.165 & $ 10 $  &5.883 & 0.663 & 3.091 & 2.131\\
    

     \hline \hline
     \end{tabular}
    \caption{Same as Table \ref{T3}, but for the $ \rm 30~M_{\odot}$ star.}
    \label{T4}
\end{sidewaystable*}






\newpage

\bibliographystyle{elsarticle-harv} 
\bibliography{example}

\begin{thebibliography}{78}
\expandafter\ifx\csname natexlab\endcsname\relax\def\natexlab#1{#1}\fi
\providecommand{\url}[1]{\texttt{#1}}
\providecommand{\href}[2]{#2}
\providecommand{\path}[1]{#1}
\providecommand{\DOIprefix}{doi:}
\providecommand{\ArXivprefix}{arXiv:}
\providecommand{\URLprefix}{URL: }
\providecommand{\Pubmedprefix}{pmid:}
\providecommand{\doi}[1]{\href{http://dx.doi.org/#1}{\path{#1}}}
\providecommand{\Pubmed}[1]{\href{pmid:#1}{\path{#1}}}
\providecommand{\bibinfo}[2]{#2}
\ifx\xfnm\relax \def\xfnm[#1]{\unskip,\space#1}\fi
\bibitem[{{Bachetti} et~al.(2014){Bachetti}, {Harrison}, {Walton}, {Grefenstette}, {Chakrabarty}, {F{\"u}rst}, {Barret}, {Beloborodov}, {Boggs}, {Christensen}, {Craig}, {Fabian}, {Hailey}, {Hornschemeier}, {Kaspi}, {Kulkarni}, {Maccarone}, {Miller}, {Rana}, {Stern}, {Tendulkar}, {Tomsick}, {Webb} and {Zhang}}]{2014Natur.514..202B}
\bibinfo{author}{{Bachetti}, M.}, \bibinfo{author}{{Harrison}, F.A.}, \bibinfo{author}{{Walton}, D.J.}, \bibinfo{author}{{Grefenstette}, B.W.}, \bibinfo{author}{{Chakrabarty}, D.}, \bibinfo{author}{{F{\"u}rst}, F.}, \bibinfo{author}{{Barret}, D.}, \bibinfo{author}{{Beloborodov}, A.}, \bibinfo{author}{{Boggs}, S.E.}, \bibinfo{author}{{Christensen}, F.E.}, \bibinfo{author}{{Craig}, W.W.}, \bibinfo{author}{{Fabian}, A.C.}, \bibinfo{author}{{Hailey}, C.J.}, \bibinfo{author}{{Hornschemeier}, A.}, \bibinfo{author}{{Kaspi}, V.}, \bibinfo{author}{{Kulkarni}, S.R.}, \bibinfo{author}{{Maccarone}, T.}, \bibinfo{author}{{Miller}, J.M.}, \bibinfo{author}{{Rana}, V.}, \bibinfo{author}{{Stern}, D.}, \bibinfo{author}{{Tendulkar}, S.P.}, \bibinfo{author}{{Tomsick}, J.}, \bibinfo{author}{{Webb}, N.A.}, \bibinfo{author}{{Zhang}, W.W.}, \bibinfo{year}{2014}.
\newblock \bibinfo{title}{{An ultraluminous X-ray source powered by an accreting neutron star}}.
\newblock \bibinfo{journal}{\nat} \bibinfo{volume}{514}, \bibinfo{pages}{202--204}.
\newblock \DOIprefix\doi{10.1038/nature13791}, \href{http://arxiv.org/abs/1410.3590}{{\tt arXiv:1410.3590}}.
\bibitem[{{Baraffe} and {El Eid}(1991)}]{1991A&A...245..548B}
\bibinfo{author}{{Baraffe}, I.}, \bibinfo{author}{{El Eid}, M.F.}, \bibinfo{year}{1991}.
\newblock \bibinfo{title}{{Evolution of massive stars with variable initial compositions}}.
\newblock \bibinfo{journal}{\aap} \bibinfo{volume}{245}, \bibinfo{pages}{548--560}.
\bibitem[{{Bear} and {Soker}(2024)}]{2024arXiv240703182B}
\bibinfo{author}{{Bear}, E.}, \bibinfo{author}{{Soker}, N.}, \bibinfo{year}{2024}.
\newblock \bibinfo{title}{{On the response of massive main sequence stars to mass accretion and outflow at high rates}}.
\newblock \bibinfo{journal}{arXiv e-prints} , \bibinfo{pages}{arXiv:2407.03182}\DOIprefix\doi{10.48550/arXiv.2407.03182}, \href{http://arxiv.org/abs/2407.03182}{{\tt arXiv:2407.03182}}.
\bibitem[{{Berger} et~al.(2009){Berger}, {Soderberg}, {Chevalier}, {Fransson}, {Foley}, {Leonard}, {Debes}, {Diamond-Stanic}, {Dupree}, {Ivans}, {Simmerer}, {Thompson} and {Tremonti}}]{2009ApJ...699.1850B}
\bibinfo{author}{{Berger}, E.}, \bibinfo{author}{{Soderberg}, A.M.}, \bibinfo{author}{{Chevalier}, R.A.}, \bibinfo{author}{{Fransson}, C.}, \bibinfo{author}{{Foley}, R.J.}, \bibinfo{author}{{Leonard}, D.C.}, \bibinfo{author}{{Debes}, J.H.}, \bibinfo{author}{{Diamond-Stanic}, A.M.}, \bibinfo{author}{{Dupree}, A.K.}, \bibinfo{author}{{Ivans}, I.I.}, \bibinfo{author}{{Simmerer}, J.}, \bibinfo{author}{{Thompson}, I.B.}, \bibinfo{author}{{Tremonti}, C.A.}, \bibinfo{year}{2009}.
\newblock \bibinfo{title}{{An Intermediate Luminosity Transient in NGC 300: The Eruption of a Dust-Enshrouded Massive Star}}.
\newblock \bibinfo{journal}{\apj} \bibinfo{volume}{699}, \bibinfo{pages}{1850--1865}.
\newblock \DOIprefix\doi{10.1088/0004-637X/699/2/1850}, \href{http://arxiv.org/abs/0901.0710}{{\tt arXiv:0901.0710}}.
\bibitem[{{Brott} et~al.(2011){Brott}, {de Mink}, {Cantiello}, {Langer}, {de Koter}, {Evans}, {Hunter}, {Trundle} and {Vink}}]{2011A&A...530A.115B}
\bibinfo{author}{{Brott}, I.}, \bibinfo{author}{{de Mink}, S.E.}, \bibinfo{author}{{Cantiello}, M.}, \bibinfo{author}{{Langer}, N.}, \bibinfo{author}{{de Koter}, A.}, \bibinfo{author}{{Evans}, C.J.}, \bibinfo{author}{{Hunter}, I.}, \bibinfo{author}{{Trundle}, C.}, \bibinfo{author}{{Vink}, J.S.}, \bibinfo{year}{2011}.
\newblock \bibinfo{title}{{Rotating massive main-sequence stars. I. Grids of evolutionary models and isochrones}}.
\newblock \bibinfo{journal}{\aap} \bibinfo{volume}{530}, \bibinfo{pages}{A115}.
\newblock \DOIprefix\doi{10.1051/0004-6361/201016113}, \href{http://arxiv.org/abs/1102.0530}{{\tt arXiv:1102.0530}}.
\bibitem[{{Brunish} and {Truran}(1982)}]{1982ApJS...49..447B}
\bibinfo{author}{{Brunish}, W.M.}, \bibinfo{author}{{Truran}, J.W.}, \bibinfo{year}{1982}.
\newblock \bibinfo{title}{{The evolution of massive stars. II. The influence of initial composition and mass loss.}}
\newblock \bibinfo{journal}{\apjs} \bibinfo{volume}{49}, \bibinfo{pages}{447--468}.
\newblock \DOIprefix\doi{10.1086/190806}.
\bibitem[{{de Mink} et~al.(2013){de Mink}, {Langer}, {Izzard}, {Sana} and {de Koter}}]{2013ApJ...764..166D}
\bibinfo{author}{{de Mink}, S.E.}, \bibinfo{author}{{Langer}, N.}, \bibinfo{author}{{Izzard}, R.G.}, \bibinfo{author}{{Sana}, H.}, \bibinfo{author}{{de Koter}, A.}, \bibinfo{year}{2013}.
\newblock \bibinfo{title}{{The Rotation Rates of Massive Stars: The Role of Binary Interaction through Tides, Mass Transfer, and Mergers}}.
\newblock \bibinfo{journal}{\apj} \bibinfo{volume}{764}, \bibinfo{pages}{166}.
\newblock \DOIprefix\doi{10.1088/0004-637X/764/2/166}, \href{http://arxiv.org/abs/1211.3742}{{\tt arXiv:1211.3742}}.
\bibitem[{{Georgy} et~al.(2013){Georgy}, {Ekstr{\"o}m}, {Eggenberger}, {Meynet}, {Haemmerl{\'e}}, {Maeder}, {Granada}, {Groh}, {Hirschi}, {Mowlavi}, {Yusof}, {Charbonnel}, {Decressin} and {Barblan}}]{2013A&A...558A.103G}
\bibinfo{author}{{Georgy}, C.}, \bibinfo{author}{{Ekstr{\"o}m}, S.}, \bibinfo{author}{{Eggenberger}, P.}, \bibinfo{author}{{Meynet}, G.}, \bibinfo{author}{{Haemmerl{\'e}}, L.}, \bibinfo{author}{{Maeder}, A.}, \bibinfo{author}{{Granada}, A.}, \bibinfo{author}{{Groh}, J.H.}, \bibinfo{author}{{Hirschi}, R.}, \bibinfo{author}{{Mowlavi}, N.}, \bibinfo{author}{{Yusof}, N.}, \bibinfo{author}{{Charbonnel}, C.}, \bibinfo{author}{{Decressin}, T.}, \bibinfo{author}{{Barblan}, F.}, \bibinfo{year}{2013}.
\newblock \bibinfo{title}{{Grids of stellar models with rotation. III. Models from 0.8 to 120 M$_{{\ensuremath{\odot}}}$ at a metallicity Z = 0.002}}.
\newblock \bibinfo{journal}{\aap} \bibinfo{volume}{558}, \bibinfo{pages}{A103}.
\newblock \DOIprefix\doi{10.1051/0004-6361/201322178}, \href{http://arxiv.org/abs/1308.2914}{{\tt arXiv:1308.2914}}.
\bibitem[{{Groh} et~al.(2019){Groh}, {Ekstr{\"o}m}, {Georgy}, {Meynet}, {Choplin}, {Eggenberger}, {Hirschi}, {Maeder}, {Murphy}, {Boian} and {Farrell}}]{2019A&A...627A..24G}
\bibinfo{author}{{Groh}, J.H.}, \bibinfo{author}{{Ekstr{\"o}m}, S.}, \bibinfo{author}{{Georgy}, C.}, \bibinfo{author}{{Meynet}, G.}, \bibinfo{author}{{Choplin}, A.}, \bibinfo{author}{{Eggenberger}, P.}, \bibinfo{author}{{Hirschi}, R.}, \bibinfo{author}{{Maeder}, A.}, \bibinfo{author}{{Murphy}, L.J.}, \bibinfo{author}{{Boian}, I.}, \bibinfo{author}{{Farrell}, E.J.}, \bibinfo{year}{2019}.
\newblock \bibinfo{title}{{Grids of stellar models with rotation. IV. Models from 1.7 to 120 M$_{{\ensuremath{\odot}}}$ at a metallicity Z = 0.0004}}.
\newblock \bibinfo{journal}{\aap} \bibinfo{volume}{627}, \bibinfo{pages}{A24}.
\newblock \DOIprefix\doi{10.1051/0004-6361/201833720}, \href{http://arxiv.org/abs/1904.04009}{{\tt arXiv:1904.04009}}.
\bibitem[{{Heger} et~al.(2000){Heger}, {Langer} and {Woosley}}]{2000ApJ...528..368H}
\bibinfo{author}{{Heger}, A.}, \bibinfo{author}{{Langer}, N.}, \bibinfo{author}{{Woosley}, S.E.}, \bibinfo{year}{2000}.
\newblock \bibinfo{title}{{Presupernova Evolution of Rotating Massive Stars. I. Numerical Method and Evolution of the Internal Stellar Structure}}.
\newblock \bibinfo{journal}{\apj} \bibinfo{volume}{528}, \bibinfo{pages}{368--396}.
\newblock \DOIprefix\doi{10.1086/308158}, \href{http://arxiv.org/abs/astro-ph/9904132}{{\tt arXiv:astro-ph/9904132}}.
\bibitem[{{Heger} et~al.(2005){Heger}, {Woosley} and {Spruit}}]{2005ApJ...626..350H}
\bibinfo{author}{{Heger}, A.}, \bibinfo{author}{{Woosley}, S.E.}, \bibinfo{author}{{Spruit}, H.C.}, \bibinfo{year}{2005}.
\newblock \bibinfo{title}{{Presupernova Evolution of Differentially Rotating Massive Stars Including Magnetic Fields}}.
\newblock \bibinfo{journal}{\apj} \bibinfo{volume}{626}, \bibinfo{pages}{350--363}.
\newblock \DOIprefix\doi{10.1086/429868}, \href{http://arxiv.org/abs/astro-ph/0409422}{{\tt arXiv:astro-ph/0409422}}.
\bibitem[{{Henyey} et~al.(1965){Henyey}, {Vardya} and {Bodenheimer}}]{1965ApJ...142..841H}
\bibinfo{author}{{Henyey}, L.}, \bibinfo{author}{{Vardya}, M.S.}, \bibinfo{author}{{Bodenheimer}, P.}, \bibinfo{year}{1965}.
\newblock \bibinfo{title}{{Studies in Stellar Evolution. III. The Calculation of Model Envelopes.}}
\newblock \bibinfo{journal}{\apj} \bibinfo{volume}{142}, \bibinfo{pages}{841}.
\newblock \DOIprefix\doi{10.1086/148357}.
\bibitem[{{Herwig}(2000)}]{2000A&A...360..952H}
\bibinfo{author}{{Herwig}, F.}, \bibinfo{year}{2000}.
\newblock \bibinfo{title}{{The evolution of AGB stars with convective overshoot}}.
\newblock \bibinfo{journal}{\aap} \bibinfo{volume}{360}, \bibinfo{pages}{952--968}.
\newblock \DOIprefix\doi{10.48550/arXiv.astro-ph/0007139}, \href{http://arxiv.org/abs/astro-ph/0007139}{{\tt arXiv:astro-ph/0007139}}.
\bibitem[{{Hjellming} and {Taam}(1991)}]{1991ApJ...370..709H}
\bibinfo{author}{{Hjellming}, M.S.}, \bibinfo{author}{{Taam}, R.E.}, \bibinfo{year}{1991}.
\newblock \bibinfo{title}{{The Response of Main-Sequence Stars within a Common Envelope}}.
\newblock \bibinfo{journal}{\apj} \bibinfo{volume}{370}, \bibinfo{pages}{709}.
\newblock \DOIprefix\doi{10.1086/169854}.
\bibitem[{{Hunt}(1971)}]{1971MNRAS.154..141H}
\bibinfo{author}{{Hunt}, R.}, \bibinfo{year}{1971}.
\newblock \bibinfo{title}{{A fluid dynamical study of the accretion process}}.
\newblock \bibinfo{journal}{\mnras} \bibinfo{volume}{154}, \bibinfo{pages}{141}.
\newblock \DOIprefix\doi{10.1093/mnras/154.2.141}.
\bibitem[{{Hurley} et~al.(2002){Hurley}, {Tout} and {Pols}}]{2002MNRAS.329..897H}
\bibinfo{author}{{Hurley}, J.R.}, \bibinfo{author}{{Tout}, C.A.}, \bibinfo{author}{{Pols}, O.R.}, \bibinfo{year}{2002}.
\newblock \bibinfo{title}{{Evolution of binary stars and the effect of tides on binary populations}}.
\newblock \bibinfo{journal}{\mnras} \bibinfo{volume}{329}, \bibinfo{pages}{897--928}.
\newblock \DOIprefix\doi{10.1046/j.1365-8711.2002.05038.x}, \href{http://arxiv.org/abs/astro-ph/0201220}{{\tt arXiv:astro-ph/0201220}}.
\bibitem[{{Iglesias} and {Rogers}(1996)}]{1996ApJ...464..943I}
\bibinfo{author}{{Iglesias}, C.A.}, \bibinfo{author}{{Rogers}, F.J.}, \bibinfo{year}{1996}.
\newblock \bibinfo{title}{{Updated Opal Opacities}}.
\newblock \bibinfo{journal}{\apj} \bibinfo{volume}{464}, \bibinfo{pages}{943}.
\newblock \DOIprefix\doi{10.1086/177381}.
\bibitem[{{Ishii} et~al.(1999){Ishii}, {Ueno} and {Kato}}]{1999PASJ...51..417I}
\bibinfo{author}{{Ishii}, M.}, \bibinfo{author}{{Ueno}, M.}, \bibinfo{author}{{Kato}, M.}, \bibinfo{year}{1999}.
\newblock \bibinfo{title}{{Core-Halo Structure of a Chemically Homogeneous Massive Star and Bending of the Zero-Age Main Sequence}}.
\newblock \bibinfo{journal}{\pasj} \bibinfo{volume}{51}, \bibinfo{pages}{417--424}.
\newblock \DOIprefix\doi{10.1093/pasj/51.4.417}, \href{http://arxiv.org/abs/astro-ph/9907154}{{\tt arXiv:astro-ph/9907154}}.
\bibitem[{{Ivanova} et~al.(2013){Ivanova}, {Justham}, {Chen}, {De Marco}, {Fryer}, {Gaburov}, {Ge}, {Glebbeek}, {Han}, {Li}, {Lu}, {Marsh}, {Podsiadlowski}, {Potter}, {Soker}, {Taam}, {Tauris}, {van den Heuvel} and {Webbink}}]{2013A&ARv..21...59I}
\bibinfo{author}{{Ivanova}, N.}, \bibinfo{author}{{Justham}, S.}, \bibinfo{author}{{Chen}, X.}, \bibinfo{author}{{De Marco}, O.}, \bibinfo{author}{{Fryer}, C.L.}, \bibinfo{author}{{Gaburov}, E.}, \bibinfo{author}{{Ge}, H.}, \bibinfo{author}{{Glebbeek}, E.}, \bibinfo{author}{{Han}, Z.}, \bibinfo{author}{{Li}, X.D.}, \bibinfo{author}{{Lu}, G.}, \bibinfo{author}{{Marsh}, T.}, \bibinfo{author}{{Podsiadlowski}, P.}, \bibinfo{author}{{Potter}, A.}, \bibinfo{author}{{Soker}, N.}, \bibinfo{author}{{Taam}, R.}, \bibinfo{author}{{Tauris}, T.M.}, \bibinfo{author}{{van den Heuvel}, E.P.J.}, \bibinfo{author}{{Webbink}, R.F.}, \bibinfo{year}{2013}.
\newblock \bibinfo{title}{{Common envelope evolution: where we stand and how we can move forward}}.
\newblock \bibinfo{journal}{\aapr} \bibinfo{volume}{21}, \bibinfo{pages}{59}.
\newblock \DOIprefix\doi{10.1007/s00159-013-0059-2}, \href{http://arxiv.org/abs/1209.4302}{{\tt arXiv:1209.4302}}.
\bibitem[{{Kashi}(2010)}]{2010MNRAS.405.1924K}
\bibinfo{author}{{Kashi}, A.}, \bibinfo{year}{2010}.
\newblock \bibinfo{title}{{An indication for the binarity of P Cygni from its 17th century eruption}}.
\newblock \bibinfo{journal}{\mnras} \bibinfo{volume}{405}, \bibinfo{pages}{1924--1929}.
\newblock \DOIprefix\doi{10.1111/j.1365-2966.2010.16582.x}, \href{http://arxiv.org/abs/0912.3998}{{\tt arXiv:0912.3998}}.
\bibitem[{{Kashi} et~al.(2010){Kashi}, {Frankowski} and {Soker}}]{2010ApJ...709L..11K}
\bibinfo{author}{{Kashi}, A.}, \bibinfo{author}{{Frankowski}, A.}, \bibinfo{author}{{Soker}, N.}, \bibinfo{year}{2010}.
\newblock \bibinfo{title}{{NGC 300 OT2008-1 as a Scaled-down Version of the Eta Carinae Great Eruption}}.
\newblock \bibinfo{journal}{\apjl} \bibinfo{volume}{709}, \bibinfo{pages}{L11--L15}.
\newblock \DOIprefix\doi{10.1088/2041-8205/709/1/L11}, \href{http://arxiv.org/abs/0909.1909}{{\tt arXiv:0909.1909}}.
\bibitem[{{Kashi} et~al.(2022){Kashi}, {Michaelis} and {Kaminetsky}}]{2022MNRAS.516.3193K}
\bibinfo{author}{{Kashi}, A.}, \bibinfo{author}{{Michaelis}, A.}, \bibinfo{author}{{Kaminetsky}, Y.}, \bibinfo{year}{2022}.
\newblock \bibinfo{title}{{Accretion in massive colliding-wind binaries and the effect of the wind momentum ratio}}.
\newblock \bibinfo{journal}{\mnras} \bibinfo{volume}{516}, \bibinfo{pages}{3193--3205}.
\newblock \DOIprefix\doi{10.1093/mnras/stac1912}, \href{http://arxiv.org/abs/2207.01990}{{\tt arXiv:2207.01990}}.
\bibitem[{{Kashi} and {Soker}(2010a)}]{2010arXiv1011.1222K}
\bibinfo{author}{{Kashi}, A.}, \bibinfo{author}{{Soker}, N.}, \bibinfo{year}{2010}a.
\newblock \bibinfo{title}{{Common Powering Mechanism of Intermediate Luminosity Optical Transients and Luminous Blue Variables}}.
\newblock \bibinfo{journal}{arXiv e-prints} , \bibinfo{pages}{arXiv:1011.1222}\DOIprefix\doi{10.48550/arXiv.1011.1222}, \href{http://arxiv.org/abs/1011.1222}{{\tt arXiv:1011.1222}}.
\bibitem[{{Kashi} and {Soker}(2010b)}]{2010ApJ...723..602K}
\bibinfo{author}{{Kashi}, A.}, \bibinfo{author}{{Soker}, N.}, \bibinfo{year}{2010}b.
\newblock \bibinfo{title}{{Periastron Passage Triggering of the 19th Century Eruptions of Eta Carinae}}.
\newblock \bibinfo{journal}{\apj} \bibinfo{volume}{723}, \bibinfo{pages}{602--611}.
\newblock \DOIprefix\doi{10.1088/0004-637X/723/1/602}, \href{http://arxiv.org/abs/0912.1439}{{\tt arXiv:0912.1439}}.
\bibitem[{{Kashi} and {Soker}(2016)}]{2016RAA....16...99K}
\bibinfo{author}{{Kashi}, A.}, \bibinfo{author}{{Soker}, N.}, \bibinfo{year}{2016}.
\newblock \bibinfo{title}{{Operation of the jet feedback mechanism (JFM) in intermediate luminosity optical transients (ILOTs)}}.
\newblock \bibinfo{journal}{Research in Astronomy and Astrophysics} \bibinfo{volume}{16}, \bibinfo{pages}{99}.
\newblock \DOIprefix\doi{10.1088/1674-4527/16/6/099}, \href{http://arxiv.org/abs/1508.00004}{{\tt arXiv:1508.00004}}.
\bibitem[{{Kashi} et~al.(2013){Kashi}, {Soker} and {Moskovitz}}]{2013MNRAS.436.2484K}
\bibinfo{author}{{Kashi}, A.}, \bibinfo{author}{{Soker}, N.}, \bibinfo{author}{{Moskovitz}, N.}, \bibinfo{year}{2013}.
\newblock \bibinfo{title}{{Powering the second 2012 outburst of SN 2009ip by repeating binary interaction}}.
\newblock \bibinfo{journal}{\mnras} \bibinfo{volume}{436}, \bibinfo{pages}{2484--2491}.
\newblock \DOIprefix\doi{10.1093/mnras/stt1742}, \href{http://arxiv.org/abs/1307.7681}{{\tt arXiv:1307.7681}}.
\bibitem[{{Kippenhahn} and {Meyer-Hofmeister}(1977)}]{1977A&A....54..539K}
\bibinfo{author}{{Kippenhahn}, R.}, \bibinfo{author}{{Meyer-Hofmeister}, E.}, \bibinfo{year}{1977}.
\newblock \bibinfo{title}{{On the radii of accreting main sequence stars.}}
\newblock \bibinfo{journal}{\aap} \bibinfo{volume}{54}, \bibinfo{pages}{539--542}.
\bibitem[{{Kippenhahn} et~al.(1980){Kippenhahn}, {Ruschenplatt} and {Thomas}}]{1980A&A....91..175K}
\bibinfo{author}{{Kippenhahn}, R.}, \bibinfo{author}{{Ruschenplatt}, G.}, \bibinfo{author}{{Thomas}, H.C.}, \bibinfo{year}{1980}.
\newblock \bibinfo{title}{{The time scale of thermohaline mixing in stars}}.
\newblock \bibinfo{journal}{\aap} \bibinfo{volume}{91}, \bibinfo{pages}{175--180}.
\bibitem[{{Kippenhahn} et~al.(2013){Kippenhahn}, {Weigert} and {Weiss}}]{2013sse..book.....K}
\bibinfo{author}{{Kippenhahn}, R.}, \bibinfo{author}{{Weigert}, A.}, \bibinfo{author}{{Weiss}, A.}, \bibinfo{year}{2013}.
\newblock \bibinfo{title}{{Stellar Structure and Evolution}}.
\newblock \DOIprefix\doi{10.1007/978-3-642-30304-3}.
\bibitem[{{Langer} et~al.(1985){Langer}, {El Eid} and {Fricke}}]{1985A&A...145..179L}
\bibinfo{author}{{Langer}, N.}, \bibinfo{author}{{El Eid}, M.F.}, \bibinfo{author}{{Fricke}, K.J.}, \bibinfo{year}{1985}.
\newblock \bibinfo{title}{{Evolution of massive stars with semiconvective diffusion}}.
\newblock \bibinfo{journal}{\aap} \bibinfo{volume}{145}, \bibinfo{pages}{179--191}.
\bibitem[{{Langer} et~al.(2003){Langer}, {Yoon}, {Petrovic} and {Heger}}]{2003astro.ph..2232L}
\bibinfo{author}{{Langer}, N.}, \bibinfo{author}{{Yoon}, S.C.}, \bibinfo{author}{{Petrovic}, J.}, \bibinfo{author}{{Heger}, A.}, \bibinfo{year}{2003}.
\newblock \bibinfo{title}{{Binary evolution models with rotation}}.
\newblock \bibinfo{journal}{arXiv e-prints} , \bibinfo{pages}{astro--ph/0302232}\DOIprefix\doi{10.48550/arXiv.astro-ph/0302232}, \href{http://arxiv.org/abs/astro-ph/0302232}{{\tt arXiv:astro-ph/0302232}}.
\bibitem[{{Lau} et~al.(2024){Lau}, {Hirai}, {Mandel} and {Tout}}]{2024ApJ...966L...7L}
\bibinfo{author}{{Lau}, M.Y.M.}, \bibinfo{author}{{Hirai}, R.}, \bibinfo{author}{{Mandel}, I.}, \bibinfo{author}{{Tout}, C.A.}, \bibinfo{year}{2024}.
\newblock \bibinfo{title}{{Expansion of Accreting Main-sequence Stars during Rapid Mass Transfer}}.
\newblock \bibinfo{journal}{\apjl} \bibinfo{volume}{966}, \bibinfo{pages}{L7}.
\newblock \DOIprefix\doi{10.3847/2041-8213/ad3d50}, \href{http://arxiv.org/abs/2401.09570}{{\tt arXiv:2401.09570}}.
\bibitem[{{Linial} and {Sari}(2017)}]{2017MNRAS.469.2441L}
\bibinfo{author}{{Linial}, I.}, \bibinfo{author}{{Sari}, R.}, \bibinfo{year}{2017}.
\newblock \bibinfo{title}{{Mass-loss through the L2 Lagrange point - application to main-sequence EMRI}}.
\newblock \bibinfo{journal}{\mnras} \bibinfo{volume}{469}, \bibinfo{pages}{2441--2454}.
\newblock \DOIprefix\doi{10.1093/mnras/stx1041}, \href{http://arxiv.org/abs/1705.01435}{{\tt arXiv:1705.01435}}.
\bibitem[{{Marchant} et~al.(2021){Marchant}, {Pappas}, {Gallegos-Garcia}, {Berry}, {Taam}, {Kalogera} and {Podsiadlowski}}]{2021A&A...650A.107M}
\bibinfo{author}{{Marchant}, P.}, \bibinfo{author}{{Pappas}, K.M.W.}, \bibinfo{author}{{Gallegos-Garcia}, M.}, \bibinfo{author}{{Berry}, C.P.L.}, \bibinfo{author}{{Taam}, R.E.}, \bibinfo{author}{{Kalogera}, V.}, \bibinfo{author}{{Podsiadlowski}, P.}, \bibinfo{year}{2021}.
\newblock \bibinfo{title}{{The role of mass transfer and common envelope evolution in the formation of merging binary black holes}}.
\newblock \bibinfo{journal}{\aap} \bibinfo{volume}{650}, \bibinfo{pages}{A107}.
\newblock \DOIprefix\doi{10.1051/0004-6361/202039992}, \href{http://arxiv.org/abs/2103.09243}{{\tt arXiv:2103.09243}}.
\bibitem[{{Marigo} et~al.(2001){Marigo}, {Girardi}, {Chiosi} and {Wood}}]{2001A&A...371..152M}
\bibinfo{author}{{Marigo}, P.}, \bibinfo{author}{{Girardi}, L.}, \bibinfo{author}{{Chiosi}, C.}, \bibinfo{author}{{Wood}, P.R.}, \bibinfo{year}{2001}.
\newblock \bibinfo{title}{{Zero-metallicity stars. I. Evolution at constant mass}}.
\newblock \bibinfo{journal}{\aap} \bibinfo{volume}{371}, \bibinfo{pages}{152--173}.
\newblock \DOIprefix\doi{10.1051/0004-6361:20010309}, \href{http://arxiv.org/abs/astro-ph/0102253}{{\tt arXiv:astro-ph/0102253}}.
\bibitem[{{Misra} et~al.(2020){Misra}, {Fragos}, {Tauris}, {Zapartas} and {Aguilera-Dena}}]{2020A&A...642A.174M}
\bibinfo{author}{{Misra}, D.}, \bibinfo{author}{{Fragos}, T.}, \bibinfo{author}{{Tauris}, T.M.}, \bibinfo{author}{{Zapartas}, E.}, \bibinfo{author}{{Aguilera-Dena}, D.R.}, \bibinfo{year}{2020}.
\newblock \bibinfo{title}{{The origin of pulsating ultra-luminous X-ray sources: Low- and intermediate-mass X-ray binaries containing neutron star accretors}}.
\newblock \bibinfo{journal}{\aap} \bibinfo{volume}{642}, \bibinfo{pages}{A174}.
\newblock \DOIprefix\doi{10.1051/0004-6361/202038070}, \href{http://arxiv.org/abs/2004.01205}{{\tt arXiv:2004.01205}}.
\bibitem[{{Mukhija} and {Kashi}(2024)}]{2024ApJ...974..124M}
\bibinfo{author}{{Mukhija}, B.}, \bibinfo{author}{{Kashi}, A.}, \bibinfo{year}{2024}.
\newblock \bibinfo{title}{{Giant Eruptions in Massive Stars and their Effect on the Stellar Structure}}.
\newblock \bibinfo{journal}{\apj} \bibinfo{volume}{974}, \bibinfo{pages}{124}.
\newblock \DOIprefix\doi{10.3847/1538-4357/ad7398}, \href{http://arxiv.org/abs/2408.01718}{{\tt arXiv:2408.01718}}.
\bibitem[{{Mukhija} and {Kashi}(2025)}]{2025ApJ...986..188M}
\bibinfo{author}{{Mukhija}, B.}, \bibinfo{author}{{Kashi}, A.}, \bibinfo{year}{2025}.
\newblock \bibinfo{title}{{Accretion and Recovery in Giant Eruptions of Massive Stars}}.
\newblock \bibinfo{journal}{\apj} \bibinfo{volume}{986}, \bibinfo{pages}{188}.
\newblock \DOIprefix\doi{10.3847/1538-4357/add3f1}, \href{http://arxiv.org/abs/2504.19884}{{\tt arXiv:2504.19884}}.
\bibitem[{{Muthukrishna} et~al.(2019){Muthukrishna}, {Narayan}, {Mandel}, {Biswas} and {Hlo{\v{z}}ek}}]{2019PASP..131k8002M}
\bibinfo{author}{{Muthukrishna}, D.}, \bibinfo{author}{{Narayan}, G.}, \bibinfo{author}{{Mandel}, K.S.}, \bibinfo{author}{{Biswas}, R.}, \bibinfo{author}{{Hlo{\v{z}}ek}, R.}, \bibinfo{year}{2019}.
\newblock \bibinfo{title}{{RAPID: Early Classification of Explosive Transients Using Deep Learning}}.
\newblock \bibinfo{journal}{\pasp} \bibinfo{volume}{131}, \bibinfo{pages}{118002}.
\newblock \DOIprefix\doi{10.1088/1538-3873/ab1609}, \href{http://arxiv.org/abs/1904.00014}{{\tt arXiv:1904.00014}}.
\bibitem[{{Nandez} et~al.(2014){Nandez}, {Ivanova} and {Lombardi}}]{2014ApJ...786...39N}
\bibinfo{author}{{Nandez}, J.L.A.}, \bibinfo{author}{{Ivanova}, N.}, \bibinfo{author}{{Lombardi}, J.~C., J.}, \bibinfo{year}{2014}.
\newblock \bibinfo{title}{{V1309 Sco{\textemdash}Understanding a Merger}}.
\newblock \bibinfo{journal}{\apj} \bibinfo{volume}{786}, \bibinfo{pages}{39}.
\newblock \DOIprefix\doi{10.1088/0004-637X/786/1/39}, \href{http://arxiv.org/abs/1311.6522}{{\tt arXiv:1311.6522}}.
\bibitem[{{Neo} et~al.(1977){Neo}, {Miyaji}, {Nomoto} and {Sugimoto}}]{1977PASJ...29..249N}
\bibinfo{author}{{Neo}, S.}, \bibinfo{author}{{Miyaji}, S.}, \bibinfo{author}{{Nomoto}, K.}, \bibinfo{author}{{Sugimoto}, D.}, \bibinfo{year}{1977}.
\newblock \bibinfo{title}{{Effect of Rapid Mass Accretion onto the Main-Sequence Stars}}.
\newblock \bibinfo{journal}{\pasj} \bibinfo{volume}{29}, \bibinfo{pages}{249--262}.
\bibitem[{{Nota} and {Lamers}(1997)}]{1997ASPC..120.....N}
\bibinfo{editor}{{Nota}, A.}, \bibinfo{editor}{{Lamers}, H.} (Eds.), \bibinfo{year}{1997}.
\newblock \bibinfo{title}{{Luminous Blue Variables: Massive Stars in Transition}}. volume \bibinfo{volume}{120} of \textit{\bibinfo{series}{Astronomical Society of the Pacific Conference Series}}.
\bibitem[{{Owocki} and {Rybicki}(1984)}]{1984ApJ...284..337O}
\bibinfo{author}{{Owocki}, S.P.}, \bibinfo{author}{{Rybicki}, G.B.}, \bibinfo{year}{1984}.
\newblock \bibinfo{title}{{Instabilities in line-driven stellar winds. I. Dependence on perturbation wavelength.}}
\newblock \bibinfo{journal}{\apj} \bibinfo{volume}{284}, \bibinfo{pages}{337--350}.
\newblock \DOIprefix\doi{10.1086/162412}.
\bibitem[{{Packet}(1981)}]{1981A&A...102...17P}
\bibinfo{author}{{Packet}, W.}, \bibinfo{year}{1981}.
\newblock \bibinfo{title}{{On the spin-up of the mass accreting component in a close binary system}}.
\newblock \bibinfo{journal}{\aap} \bibinfo{volume}{102}, \bibinfo{pages}{17--19}.
\bibitem[{{Paczy{\'n}ski}(1971)}]{1971ARA&A...9..183P}
\bibinfo{author}{{Paczy{\'n}ski}, B.}, \bibinfo{year}{1971}.
\newblock \bibinfo{title}{{Evolutionary Processes in Close Binary Systems}}.
\newblock \bibinfo{journal}{\araa} \bibinfo{volume}{9}, \bibinfo{pages}{183}.
\newblock \DOIprefix\doi{10.1146/annurev.aa.09.090171.001151}.
\bibitem[{{Paxton} et~al.(2011){Paxton}, {Bildsten}, {Dotter}, {Herwig}, {Lesaffre} and {Timmes}}]{2011ApJS..192....3P}
\bibinfo{author}{{Paxton}, B.}, \bibinfo{author}{{Bildsten}, L.}, \bibinfo{author}{{Dotter}, A.}, \bibinfo{author}{{Herwig}, F.}, \bibinfo{author}{{Lesaffre}, P.}, \bibinfo{author}{{Timmes}, F.}, \bibinfo{year}{2011}.
\newblock \bibinfo{title}{{Modules for Experiments in Stellar Astrophysics (MESA)}}.
\newblock \bibinfo{journal}{\apjs} \bibinfo{volume}{192}, \bibinfo{pages}{3}.
\newblock \DOIprefix\doi{10.1088/0067-0049/192/1/3}, \href{http://arxiv.org/abs/1009.1622}{{\tt arXiv:1009.1622}}.
\bibitem[{{Paxton} et~al.(2013){Paxton}, {Cantiello}, {Arras}, {Bildsten}, {Brown}, {Dotter}, {Mankovich}, {Montgomery}, {Stello}, {Timmes} and {Townsend}}]{2013ApJS..208....4P}
\bibinfo{author}{{Paxton}, B.}, \bibinfo{author}{{Cantiello}, M.}, \bibinfo{author}{{Arras}, P.}, \bibinfo{author}{{Bildsten}, L.}, \bibinfo{author}{{Brown}, E.F.}, \bibinfo{author}{{Dotter}, A.}, \bibinfo{author}{{Mankovich}, C.}, \bibinfo{author}{{Montgomery}, M.H.}, \bibinfo{author}{{Stello}, D.}, \bibinfo{author}{{Timmes}, F.X.}, \bibinfo{author}{{Townsend}, R.}, \bibinfo{year}{2013}.
\newblock \bibinfo{title}{{Modules for Experiments in Stellar Astrophysics (MESA): Planets, Oscillations, Rotation, and Massive Stars}}.
\newblock \bibinfo{journal}{\apjs} \bibinfo{volume}{208}, \bibinfo{pages}{4}.
\newblock \DOIprefix\doi{10.1088/0067-0049/208/1/4}, \href{http://arxiv.org/abs/1301.0319}{{\tt arXiv:1301.0319}}.
\bibitem[{{Paxton} et~al.(2015){Paxton}, {Marchant}, {Schwab}, {Bauer}, {Bildsten}, {Cantiello}, {Dessart}, {Farmer}, {Hu}, {Langer}, {Townsend}, {Townsley} and {Timmes}}]{2015ApJS..220...15P}
\bibinfo{author}{{Paxton}, B.}, \bibinfo{author}{{Marchant}, P.}, \bibinfo{author}{{Schwab}, J.}, \bibinfo{author}{{Bauer}, E.B.}, \bibinfo{author}{{Bildsten}, L.}, \bibinfo{author}{{Cantiello}, M.}, \bibinfo{author}{{Dessart}, L.}, \bibinfo{author}{{Farmer}, R.}, \bibinfo{author}{{Hu}, H.}, \bibinfo{author}{{Langer}, N.}, \bibinfo{author}{{Townsend}, R.H.D.}, \bibinfo{author}{{Townsley}, D.M.}, \bibinfo{author}{{Timmes}, F.X.}, \bibinfo{year}{2015}.
\newblock \bibinfo{title}{{Modules for Experiments in Stellar Astrophysics (MESA): Binaries, Pulsations, and Explosions}}.
\newblock \bibinfo{journal}{\apjs} \bibinfo{volume}{220}, \bibinfo{pages}{15}.
\newblock \DOIprefix\doi{10.1088/0067-0049/220/1/15}, \href{http://arxiv.org/abs/1506.03146}{{\tt arXiv:1506.03146}}.
\bibitem[{{Paxton} et~al.(2018){Paxton}, {Schwab}, {Bauer}, {Bildsten}, {Blinnikov}, {Duffell}, {Farmer}, {Goldberg}, {Marchant}, {Sorokina}, {Thoul}, {Townsend} and {Timmes}}]{2018ApJS..234...34P}
\bibinfo{author}{{Paxton}, B.}, \bibinfo{author}{{Schwab}, J.}, \bibinfo{author}{{Bauer}, E.B.}, \bibinfo{author}{{Bildsten}, L.}, \bibinfo{author}{{Blinnikov}, S.}, \bibinfo{author}{{Duffell}, P.}, \bibinfo{author}{{Farmer}, R.}, \bibinfo{author}{{Goldberg}, J.A.}, \bibinfo{author}{{Marchant}, P.}, \bibinfo{author}{{Sorokina}, E.}, \bibinfo{author}{{Thoul}, A.}, \bibinfo{author}{{Townsend}, R.H.D.}, \bibinfo{author}{{Timmes}, F.X.}, \bibinfo{year}{2018}.
\newblock \bibinfo{title}{{Modules for Experiments in Stellar Astrophysics (MESA): Convective Boundaries, Element Diffusion, and Massive Star Explosions}}.
\newblock \bibinfo{journal}{\apjs} \bibinfo{volume}{234}, \bibinfo{pages}{34}.
\newblock \DOIprefix\doi{10.3847/1538-4365/aaa5a8}, \href{http://arxiv.org/abs/1710.08424}{{\tt arXiv:1710.08424}}.
\bibitem[{{Paxton} et~al.(2019){Paxton}, {Smolec}, {Schwab}, {Gautschy}, {Bildsten}, {Cantiello}, {Dotter}, {Farmer}, {Goldberg}, {Jermyn}, {Kanbur}, {Marchant}, {Thoul}, {Townsend}, {Wolf}, {Zhang} and {Timmes}}]{2019ApJS..243...10P}
\bibinfo{author}{{Paxton}, B.}, \bibinfo{author}{{Smolec}, R.}, \bibinfo{author}{{Schwab}, J.}, \bibinfo{author}{{Gautschy}, A.}, \bibinfo{author}{{Bildsten}, L.}, \bibinfo{author}{{Cantiello}, M.}, \bibinfo{author}{{Dotter}, A.}, \bibinfo{author}{{Farmer}, R.}, \bibinfo{author}{{Goldberg}, J.A.}, \bibinfo{author}{{Jermyn}, A.S.}, \bibinfo{author}{{Kanbur}, S.M.}, \bibinfo{author}{{Marchant}, P.}, \bibinfo{author}{{Thoul}, A.}, \bibinfo{author}{{Townsend}, R.H.D.}, \bibinfo{author}{{Wolf}, W.M.}, \bibinfo{author}{{Zhang}, M.}, \bibinfo{author}{{Timmes}, F.X.}, \bibinfo{year}{2019}.
\newblock \bibinfo{title}{{Modules for Experiments in Stellar Astrophysics (MESA): Pulsating Variable Stars, Rotation, Convective Boundaries, and Energy Conservation}}.
\newblock \bibinfo{journal}{\apjs} \bibinfo{volume}{243}, \bibinfo{pages}{10}.
\newblock \DOIprefix\doi{10.3847/1538-4365/ab2241}, \href{http://arxiv.org/abs/1903.01426}{{\tt arXiv:1903.01426}}.
\bibitem[{{Petrovic} et~al.(2005){Petrovic}, {Langer} and {van der Hucht}}]{2005A&A...435.1013P}
\bibinfo{author}{{Petrovic}, J.}, \bibinfo{author}{{Langer}, N.}, \bibinfo{author}{{van der Hucht}, K.A.}, \bibinfo{year}{2005}.
\newblock \bibinfo{title}{{Constraining the mass transfer in massive binaries through progenitor evolution models of Wolf-Rayet+O binaries}}.
\newblock \bibinfo{journal}{\aap} \bibinfo{volume}{435}, \bibinfo{pages}{1013--1030}.
\newblock \DOIprefix\doi{10.1051/0004-6361:20042368}, \href{http://arxiv.org/abs/astro-ph/0504242}{{\tt arXiv:astro-ph/0504242}}.
\bibitem[{{Pinto} and {Walton}(2023)}]{2023arXiv230200006P}
\bibinfo{author}{{Pinto}, C.}, \bibinfo{author}{{Walton}, D.J.}, \bibinfo{year}{2023}.
\newblock \bibinfo{title}{{Ultra-luminous X-ray sources: extreme accretion and feedback}}.
\newblock \bibinfo{journal}{arXiv e-prints} , \bibinfo{pages}{arXiv:2302.00006}\DOIprefix\doi{10.48550/arXiv.2302.00006}, \href{http://arxiv.org/abs/2302.00006}{{\tt arXiv:2302.00006}}.
\bibitem[{{Pols} et~al.(1991){Pols}, {Cote}, {Waters} and {Heise}}]{1991A&A...241..419P}
\bibinfo{author}{{Pols}, O.R.}, \bibinfo{author}{{Cote}, J.}, \bibinfo{author}{{Waters}, L.B.F.M.}, \bibinfo{author}{{Heise}, J.}, \bibinfo{year}{1991}.
\newblock \bibinfo{title}{{The formation of Be stars through close binary evolution.}}
\newblock \bibinfo{journal}{\aap} \bibinfo{volume}{241}, \bibinfo{pages}{419}.
\bibitem[{{Pols} et~al.(2009){Pols}, {Schroeder}, {Hurley}, {Tout} and {Eggleton}}]{2009yCat..72980525P}
\bibinfo{author}{{Pols}, O.R.}, \bibinfo{author}{{Schroeder}, K.P.}, \bibinfo{author}{{Hurley}, J.R.}, \bibinfo{author}{{Tout}, C.A.}, \bibinfo{author}{{Eggleton}, P.P.}, \bibinfo{year}{2009}.
\newblock \bibinfo{title}{{VizieR Online Data Catalog: Stellar evolution models for Z = 0.0001 to 0.03 (Pols+ 1998)}}.
\newblock \bibinfo{howpublished}{VizieR On-line Data Catalog: J/MNRAS/298/525. Originally published in: 1998MNRAS.298..525P}.
\bibitem[{{Portegies Zwart} and {Verbunt}(1996)}]{1996A&A...309..179P}
\bibinfo{author}{{Portegies Zwart}, S.F.}, \bibinfo{author}{{Verbunt}, F.}, \bibinfo{year}{1996}.
\newblock \bibinfo{title}{{Population synthesis of high-mass binaries.}}
\newblock \bibinfo{journal}{\aap} \bibinfo{volume}{309}, \bibinfo{pages}{179--196}.
\bibitem[{{Pringle}(1981)}]{1981ARA&A..19..137P}
\bibinfo{author}{{Pringle}, J.E.}, \bibinfo{year}{1981}.
\newblock \bibinfo{title}{{Accretion discs in astrophysics}}.
\newblock \bibinfo{journal}{\araa} \bibinfo{volume}{19}, \bibinfo{pages}{137--162}.
\newblock \DOIprefix\doi{10.1146/annurev.aa.19.090181.001033}.
\bibitem[{{Qin} et~al.(2019){Qin}, {Marchant}, {Fragos}, {Meynet} and {Kalogera}}]{2019ApJ...870L..18Q}
\bibinfo{author}{{Qin}, Y.}, \bibinfo{author}{{Marchant}, P.}, \bibinfo{author}{{Fragos}, T.}, \bibinfo{author}{{Meynet}, G.}, \bibinfo{author}{{Kalogera}, V.}, \bibinfo{year}{2019}.
\newblock \bibinfo{title}{{On the Origin of Black Hole Spin in High-mass X-Ray Binaries}}.
\newblock \bibinfo{journal}{\apjl} \bibinfo{volume}{870}, \bibinfo{pages}{L18}.
\newblock \DOIprefix\doi{10.3847/2041-8213/aaf97b}, \href{http://arxiv.org/abs/1810.13016}{{\tt arXiv:1810.13016}}.
\bibitem[{{Quataert} et~al.(2016){Quataert}, {Fern{\'a}ndez}, {Kasen}, {Klion} and {Paxton}}]{2016MNRAS.458.1214Q}
\bibinfo{author}{{Quataert}, E.}, \bibinfo{author}{{Fern{\'a}ndez}, R.}, \bibinfo{author}{{Kasen}, D.}, \bibinfo{author}{{Klion}, H.}, \bibinfo{author}{{Paxton}, B.}, \bibinfo{year}{2016}.
\newblock \bibinfo{title}{{Super-Eddington stellar winds driven by near-surface energy deposition}}.
\newblock \bibinfo{journal}{\mnras} \bibinfo{volume}{458}, \bibinfo{pages}{1214--1233}.
\newblock \DOIprefix\doi{10.1093/mnras/stw365}, \href{http://arxiv.org/abs/1509.06370}{{\tt arXiv:1509.06370}}.
\bibitem[{{Renzo} and {G{\"o}tberg}(2021)}]{2021ApJ...923..277R}
\bibinfo{author}{{Renzo}, M.}, \bibinfo{author}{{G{\"o}tberg}, Y.}, \bibinfo{year}{2021}.
\newblock \bibinfo{title}{{Evolution of Accretor Stars in Massive Binaries: Broader Implications from Modeling {\ensuremath{\zeta}} Ophiuchi}}.
\newblock \bibinfo{journal}{\apj} \bibinfo{volume}{923}, \bibinfo{pages}{277}.
\newblock \DOIprefix\doi{10.3847/1538-4357/ac29c5}, \href{http://arxiv.org/abs/2107.10933}{{\tt arXiv:2107.10933}}.
\bibitem[{{Sanyal} et~al.(2015){Sanyal}, {Grassitelli}, {Langer} and {Bestenlehner}}]{2015A&A...580A..20S}
\bibinfo{author}{{Sanyal}, D.}, \bibinfo{author}{{Grassitelli}, L.}, \bibinfo{author}{{Langer}, N.}, \bibinfo{author}{{Bestenlehner}, J.M.}, \bibinfo{year}{2015}.
\newblock \bibinfo{title}{{Massive main-sequence stars evolving at the Eddington limit}}.
\newblock \bibinfo{journal}{\aap} \bibinfo{volume}{580}, \bibinfo{pages}{A20}.
\newblock \DOIprefix\doi{10.1051/0004-6361/201525945}, \href{http://arxiv.org/abs/1506.02997}{{\tt arXiv:1506.02997}}.
\bibitem[{{Sanyal} et~al.(2017){Sanyal}, {Langer}, {Sz{\'e}csi}, {-C Yoon} and {Grassitelli}}]{2017A&A...597A..71S}
\bibinfo{author}{{Sanyal}, D.}, \bibinfo{author}{{Langer}, N.}, \bibinfo{author}{{Sz{\'e}csi}, D.}, \bibinfo{author}{{-C Yoon}, S.}, \bibinfo{author}{{Grassitelli}, L.}, \bibinfo{year}{2017}.
\newblock \bibinfo{title}{{Metallicity dependence of envelope inflation in massive stars}}.
\newblock \bibinfo{journal}{\aap} \bibinfo{volume}{597}, \bibinfo{pages}{A71}.
\newblock \DOIprefix\doi{10.1051/0004-6361/201629612}, \href{http://arxiv.org/abs/1611.07280}{{\tt arXiv:1611.07280}}.
\bibitem[{{Schootemeijer} et~al.(2019){Schootemeijer}, {Langer}, {Grin} and {Wang}}]{2019A&A...625A.132S}
\bibinfo{author}{{Schootemeijer}, A.}, \bibinfo{author}{{Langer}, N.}, \bibinfo{author}{{Grin}, N.J.}, \bibinfo{author}{{Wang}, C.}, \bibinfo{year}{2019}.
\newblock \bibinfo{title}{{Constraining mixing in massive stars in the Small Magellanic Cloud}}.
\newblock \bibinfo{journal}{\aap} \bibinfo{volume}{625}, \bibinfo{pages}{A132}.
\newblock \DOIprefix\doi{10.1051/0004-6361/201935046}, \href{http://arxiv.org/abs/1903.10423}{{\tt arXiv:1903.10423}}.
\bibitem[{{Sch{\"u}rmann} and {Langer}(2024)}]{2024arXiv240408615S}
\bibinfo{author}{{Sch{\"u}rmann}, C.}, \bibinfo{author}{{Langer}, N.}, \bibinfo{year}{2024}.
\newblock \bibinfo{title}{{Exploring the borderline between stable mass transfer and mergers in close binary evolution}}.
\newblock \bibinfo{journal}{arXiv e-prints} , \bibinfo{pages}{arXiv:2404.08615}\DOIprefix\doi{10.48550/arXiv.2404.08615}, \href{http://arxiv.org/abs/2404.08615}{{\tt arXiv:2404.08615}}.
\bibitem[{{Scolnic} et~al.(2025){Scolnic}, {Bear} and {Soker}}]{Scolnic_2025}
\bibinfo{author}{{Scolnic}, A.}, \bibinfo{author}{{Bear}, E.}, \bibinfo{author}{{Soker}, N.}, \bibinfo{year}{2025}.
\newblock \bibinfo{title}{Enabling high mass accretion rates onto massive main sequence stars by outer envelope mass removal}.
\newblock \bibinfo{journal}{Open Journal of Astrophysics} .
\bibitem[{{Sen} et~al.(2022){Sen}, {Langer}, {Marchant}, {Menon}, {de Mink}, {Schootemeijer}, {Sch{\"u}rmann}, {Mahy}, {Hastings}, {Nathaniel}, {Sana}, {Wang} and {Xu}}]{2022A&A...659A..98S}
\bibinfo{author}{{Sen}, K.}, \bibinfo{author}{{Langer}, N.}, \bibinfo{author}{{Marchant}, P.}, \bibinfo{author}{{Menon}, A.}, \bibinfo{author}{{de Mink}, S.E.}, \bibinfo{author}{{Schootemeijer}, A.}, \bibinfo{author}{{Sch{\"u}rmann}, C.}, \bibinfo{author}{{Mahy}, L.}, \bibinfo{author}{{Hastings}, B.}, \bibinfo{author}{{Nathaniel}, K.}, \bibinfo{author}{{Sana}, H.}, \bibinfo{author}{{Wang}, C.}, \bibinfo{author}{{Xu}, X.T.}, \bibinfo{year}{2022}.
\newblock \bibinfo{title}{{Detailed models of interacting short-period massive binary stars}}.
\newblock \bibinfo{journal}{\aap} \bibinfo{volume}{659}, \bibinfo{pages}{A98}.
\newblock \DOIprefix\doi{10.1051/0004-6361/202142574}, \href{http://arxiv.org/abs/2111.03329}{{\tt arXiv:2111.03329}}.
\bibitem[{{Shao} and {Li}(2014)}]{2014ApJ...796...37S}
\bibinfo{author}{{Shao}, Y.}, \bibinfo{author}{{Li}, X.D.}, \bibinfo{year}{2014}.
\newblock \bibinfo{title}{{On the Formation of Be Stars through Binary Interaction}}.
\newblock \bibinfo{journal}{\apj} \bibinfo{volume}{796}, \bibinfo{pages}{37}.
\newblock \DOIprefix\doi{10.1088/0004-637X/796/1/37}, \href{http://arxiv.org/abs/1410.0100}{{\tt arXiv:1410.0100}}.
\bibitem[{{Shao} and {Li}(2016)}]{2016ApJ...833..108S}
\bibinfo{author}{{Shao}, Y.}, \bibinfo{author}{{Li}, X.D.}, \bibinfo{year}{2016}.
\newblock \bibinfo{title}{{Nonconservative Mass Transfer in Massive Binaries and the Formation of Wolf-Rayet+O Binaries}}.
\newblock \bibinfo{journal}{\apj} \bibinfo{volume}{833}, \bibinfo{pages}{108}.
\newblock \DOIprefix\doi{10.3847/1538-4357/833/1/108}, \href{http://arxiv.org/abs/1610.04307}{{\tt arXiv:1610.04307}}.
\bibitem[{{Shiber} et~al.(2016){Shiber}, {Schreier} and {Soker}}]{2016RAA....16..117S}
\bibinfo{author}{{Shiber}, S.}, \bibinfo{author}{{Schreier}, R.}, \bibinfo{author}{{Soker}, N.}, \bibinfo{year}{2016}.
\newblock \bibinfo{title}{{Binary interactions with high accretion rates onto main sequence stars}}.
\newblock \bibinfo{journal}{Research in Astronomy and Astrophysics} \bibinfo{volume}{16}, \bibinfo{pages}{117}.
\newblock \DOIprefix\doi{10.1088/1674-4527/16/7/117}, \href{http://arxiv.org/abs/1504.04144}{{\tt arXiv:1504.04144}}.
\bibitem[{{Soker}(2001)}]{2001MNRAS.325..584S}
\bibinfo{author}{{Soker}, N.}, \bibinfo{year}{2001}.
\newblock \bibinfo{title}{{The departure of {\ensuremath{\eta}} Carinae from axisymmetry and the binary hypothesis}}.
\newblock \bibinfo{journal}{\mnras} \bibinfo{volume}{325}, \bibinfo{pages}{584--588}.
\newblock \DOIprefix\doi{10.1046/j.1365-8711.2001.04439.x}, \href{http://arxiv.org/abs/astro-ph/0103033}{{\tt arXiv:astro-ph/0103033}}.
\bibitem[{{Soker}(2017)}]{2017MNRAS.471.4839S}
\bibinfo{author}{{Soker}, N.}, \bibinfo{year}{2017}.
\newblock \bibinfo{title}{{Energizing the last phase of common-envelope removal}}.
\newblock \bibinfo{journal}{\mnras} \bibinfo{volume}{471}, \bibinfo{pages}{4839--4843}.
\newblock \DOIprefix\doi{10.1093/mnras/stx1978}, \href{http://arxiv.org/abs/1706.03720}{{\tt arXiv:1706.03720}}.
\bibitem[{{Soker}(2020)}]{2020Galax...8...26S}
\bibinfo{author}{{Soker}, N.}, \bibinfo{year}{2020}.
\newblock \bibinfo{title}{{Shaping Planetary Nebulae with Jets and the Grazing Envelope Evolution}}.
\newblock \bibinfo{journal}{Galaxies} \bibinfo{volume}{8}, \bibinfo{pages}{26}.
\newblock \DOIprefix\doi{10.3390/galaxies8010026}, \href{http://arxiv.org/abs/2002.04229}{{\tt arXiv:2002.04229}}.
\bibitem[{{Soker}(2023)}]{2023MNRAS.524L..94S}
\bibinfo{author}{{Soker}, N.}, \bibinfo{year}{2023}.
\newblock \bibinfo{title}{{On the nature of the planet-powered transient event ZTF SLRN-2020}}.
\newblock \bibinfo{journal}{\mnras} \bibinfo{volume}{524}, \bibinfo{pages}{L94--L97}.
\newblock \DOIprefix\doi{10.1093/mnrasl/slad086}, \href{http://arxiv.org/abs/2305.04909}{{\tt arXiv:2305.04909}}.
\bibitem[{{Temmink} et~al.(2023){Temmink}, {Pols}, {Justham}, {Istrate} and {Toonen}}]{2023A&A...669A..45T}
\bibinfo{author}{{Temmink}, K.D.}, \bibinfo{author}{{Pols}, O.R.}, \bibinfo{author}{{Justham}, S.}, \bibinfo{author}{{Istrate}, A.G.}, \bibinfo{author}{{Toonen}, S.}, \bibinfo{year}{2023}.
\newblock \bibinfo{title}{{Coping with loss. Stability of mass transfer from post-main-sequence donor stars}}.
\newblock \bibinfo{journal}{\aap} \bibinfo{volume}{669}, \bibinfo{pages}{A45}.
\newblock \DOIprefix\doi{10.1051/0004-6361/202244137}, \href{http://arxiv.org/abs/2209.12707}{{\tt arXiv:2209.12707}}.
\bibitem[{{Toonen} et~al.(2012){Toonen}, {Nelemans} and {Portegies Zwart}}]{2012A&A...546A..70T}
\bibinfo{author}{{Toonen}, S.}, \bibinfo{author}{{Nelemans}, G.}, \bibinfo{author}{{Portegies Zwart}, S.}, \bibinfo{year}{2012}.
\newblock \bibinfo{title}{{Supernova Type Ia progenitors from merging double white dwarfs. Using a new population synthesis model}}.
\newblock \bibinfo{journal}{\aap} \bibinfo{volume}{546}, \bibinfo{pages}{A70}.
\newblock \DOIprefix\doi{10.1051/0004-6361/201218966}, \href{http://arxiv.org/abs/1208.6446}{{\tt arXiv:1208.6446}}.
\bibitem[{{Tylenda} et~al.(2011){Tylenda}, {Hajduk}, {Kami{\'n}ski}, {Udalski}, {Soszy{\'n}ski}, {Szyma{\'n}ski}, {Kubiak}, {Pietrzy{\'n}ski}, {Poleski}, {Wyrzykowski} and {Ulaczyk}}]{2011A&A...528A.114T}
\bibinfo{author}{{Tylenda}, R.}, \bibinfo{author}{{Hajduk}, M.}, \bibinfo{author}{{Kami{\'n}ski}, T.}, \bibinfo{author}{{Udalski}, A.}, \bibinfo{author}{{Soszy{\'n}ski}, I.}, \bibinfo{author}{{Szyma{\'n}ski}, M.K.}, \bibinfo{author}{{Kubiak}, M.}, \bibinfo{author}{{Pietrzy{\'n}ski}, G.}, \bibinfo{author}{{Poleski}, R.}, \bibinfo{author}{{Wyrzykowski}, {\L}.}, \bibinfo{author}{{Ulaczyk}, K.}, \bibinfo{year}{2011}.
\newblock \bibinfo{title}{{V1309 Scorpii: merger of a contact binary}}.
\newblock \bibinfo{journal}{\aap} \bibinfo{volume}{528}, \bibinfo{pages}{A114}.
\newblock \DOIprefix\doi{10.1051/0004-6361/201016221}, \href{http://arxiv.org/abs/1012.0163}{{\tt arXiv:1012.0163}}.
\bibitem[{{van Son} et~al.(2022){van Son}, {de Mink}, {Callister}, {Justham}, {Renzo}, {Wagg}, {Broekgaarden}, {Kummer}, {Pakmor} and {Mandel}}]{2022ApJ...931...17V}
\bibinfo{author}{{van Son}, L.A.C.}, \bibinfo{author}{{de Mink}, S.E.}, \bibinfo{author}{{Callister}, T.}, \bibinfo{author}{{Justham}, S.}, \bibinfo{author}{{Renzo}, M.}, \bibinfo{author}{{Wagg}, T.}, \bibinfo{author}{{Broekgaarden}, F.S.}, \bibinfo{author}{{Kummer}, F.}, \bibinfo{author}{{Pakmor}, R.}, \bibinfo{author}{{Mandel}, I.}, \bibinfo{year}{2022}.
\newblock \bibinfo{title}{{The Redshift Evolution of the Binary Black Hole Merger Rate: A Weighty Matter}}.
\newblock \bibinfo{journal}{\apj} \bibinfo{volume}{931}, \bibinfo{pages}{17}.
\newblock \DOIprefix\doi{10.3847/1538-4357/ac64a3}, \href{http://arxiv.org/abs/2110.01634}{{\tt arXiv:2110.01634}}.
\bibitem[{{Vigna-G{\'o}mez} et~al.(2018){Vigna-G{\'o}mez}, {Neijssel}, {Stevenson}, {Barrett}, {Belczynski}, {Justham}, {de Mink}, {M{\"u}ller}, {Podsiadlowski}, {Renzo}, {Sz{\'e}csi} and {Mandel}}]{2018MNRAS.481.4009V}
\bibinfo{author}{{Vigna-G{\'o}mez}, A.}, \bibinfo{author}{{Neijssel}, C.J.}, \bibinfo{author}{{Stevenson}, S.}, \bibinfo{author}{{Barrett}, J.W.}, \bibinfo{author}{{Belczynski}, K.}, \bibinfo{author}{{Justham}, S.}, \bibinfo{author}{{de Mink}, S.E.}, \bibinfo{author}{{M{\"u}ller}, B.}, \bibinfo{author}{{Podsiadlowski}, P.}, \bibinfo{author}{{Renzo}, M.}, \bibinfo{author}{{Sz{\'e}csi}, D.}, \bibinfo{author}{{Mandel}, I.}, \bibinfo{year}{2018}.
\newblock \bibinfo{title}{{On the formation history of Galactic double neutron stars}}.
\newblock \bibinfo{journal}{\mnras} \bibinfo{volume}{481}, \bibinfo{pages}{4009--4029}.
\newblock \DOIprefix\doi{10.1093/mnras/sty2463}, \href{http://arxiv.org/abs/1805.07974}{{\tt arXiv:1805.07974}}.
\bibitem[{{Zhao} and {Fuller}(2020)}]{2020MNRAS.495..249Z}
\bibinfo{author}{{Zhao}, X.}, \bibinfo{author}{{Fuller}, J.}, \bibinfo{year}{2020}.
\newblock \bibinfo{title}{{Centrifugally driven mass-loss and outbursts of massive stars}}.
\newblock \bibinfo{journal}{\mnras} \bibinfo{volume}{495}, \bibinfo{pages}{249--265}.
\newblock \DOIprefix\doi{10.1093/mnras/staa1097}, \href{http://arxiv.org/abs/2004.07279}{{\tt arXiv:2004.07279}}.

\end{thebibliography}

\end{document}